\documentclass[11pt]{llncs} 
\usepackage[latin1]{inputenc}
\usepackage{amsmath}
\usepackage{amsfonts}
\usepackage{amssymb}
\usepackage{graphicx}
\usepackage[Algorithme]{algorithm}
\usepackage{algorithmic}      
\usepackage{fancybox}
\usepackage{multicol}
\usepackage{color}
\usepackage{algorithmic}
\usepackage{algorithm}
\usepackage{fullpage}

\usepackage{xspace}

\newcommand{\PclassX}{{\sf\bf P}}
\newcommand{\Pclass}{\PclassX\xspace}
\newcommand{\NPclassX}{{\sf\bf NP}}
\newcommand{\NPclass}{\NPclassX\xspace}

\newcommand{\APXclassX}{{\sf\bf APX}}

\newcommand{\complete}{%
        \text{--{complete}}}

\newcommand{\NPcomplete}{\NPclassX\complete\xspace}
\newcommand{\NPC}{\NPcomplete}

\newcommand{\APXcomplete}{\APXclassX\complete\xspace}
\newcommand{\APXC}{\APXcomplete}

\newcommand{\hard}{%
        \text{--{hard}}}

\newcommand{\APXhard}{\APXclassX\hard\xspace}

\newcommand{\hardness}{%
        \text{--{hardness}}}

\newcommand{\APXhardness}{\APXclassX\hardness\xspace}

\newcommand{\PB}[1]{\textsf{\scshape{#1}}}
\newcommand{\OPTname}{opt}
\newcommand{\OPT}{\text{$\mathsf{\OPTname}$}}

\newcommand{\ALGO}[1]{\textnormal{\ttfamily\sf #1}}

\floatname{algorithm}{Algorithm}

\newcommand{\mabox}[2][0.90]
{
  \fbox{
    \begin{minipage}[c]{#1\linewidth} 
      #2  
    \end{minipage}
  }~\\
}

\newcommand{\OCC}{\textnormal{occ}}
\newcommand{\GP}{\texttt{g}}

\newtheorem{observation}[theorem]{Observation}

\title{On the Approximability of Comparing Genomes with Duplicates}

\begin{document}

\pagestyle{plain}

\author{%
  S\'ebastien Angibaud$^1$
  \and 
  Guillaume Fertin$^1$
  \and
  Irena Rusu$^1$
  \and
  Annelyse Th\'evenin$^2$
  \and 
  St\'ephane Vialette$^3$
}

\institute{$^1$ Laboratoire d'Informatique de
  Nantes-Atlantique (LINA), UMR CNRS 6241,
  Universit\'e de Nantes, 2 rue de la Houssini\`ere, 44322 Nantes
  Cedex 3, France\\ 
  \texttt{\{sebastien.angibaud,guillaume.fertin,irena.rusu\}@univ-nantes.fr}\\
  $^2$Laboratoire de Recherche en Informatique
  (LRI), UMR CNRS 8623,
  Universit\'e Paris-Sud, 91405 Orsay, France\\
  \texttt{thevenin@lri.fr}\\
  $^3$ IGM-LabInfo, UMR CNRS 8049, Universit\'e
  Paris-Est, 5 Bd Descartes 77454 Marne-la-Vall\'ee, France\\ 
  \texttt{vialette@univ-mlv.fr}}
\maketitle

\begin{abstract}
  A central problem in comparative genomics consists in computing a
  (dis-)simi\-la\-ri\-ty measure between two genomes, e.g. in order to
  construct a phylogenetic tree. A large number of such measures has
  been proposed in the recent past: \emph{number of reversals},
  \emph{number of breakpoints}, \emph{number of common} or
  \emph{conserved intervals}, \emph{SAD} etc. In their initial
  definitions, all these measures suppose that genomes contain no
  duplicates. However, we now know that genes can be duplicated within
  the same genome. One possible approach to overcome this difficulty is
  to establish a one-to-one correspondence (i.e. a matching) between
  genes of both genomes, where the correspondence is chosen in order to
  optimize the studied measure. Then, after a gene relabeling according
  to this matching and a deletion of the unmatched signed genes, two
  genomes without duplicates are obtained and the measure can be
  computed.

  In this paper, we are interested in three measures (\emph{number of
    breakpoints}, \emph{number of common intervals} and \emph{number of
    conserved intervals}) and three models of matching (\emph{exemplar},
  \emph{intermediate} and \emph{maximum matching} 
  models). 
  We prove that, for each model and each measure $\mathcal{M}$, computing a 
  matching between two genomes that optimizes $\mathcal{M}$ is 
  \APXhard. We show that this result remains true even for two
  genomes $G_1$ and $G_2$ such that $G_1$ contains no duplicates and no
  gene of $G_2$ appears more than twice. Therefore, our results extend
  those of~\cite{Complexite_exemplarisation,NP_completude_EComI_MComI,chen_inproceding_approximation_breakpoint}.
  Besides, in order to evaluate the possible existence of approximation
  algorithms concerning the number of
  breakpoints, we also study the complexity of the
  following decision problem: is there an exemplarization (resp. an
  intermediate matching, a maximum matching) that induces no
  breakpoint~? In particular, we extend a result
  of~\cite{chen_inproceding_approximation_breakpoint} by proving the
  problem to be \NPC in the exemplar model for a new class of
  instances,  we note that the problems are equivalent in the
  intermediate and the exemplar models and we show that the problem is in \Pclass in the
  maximum matching model.
  Finally, we focus on a fourth measure, closely related to the number of breakpoints: the \emph{number
    of  adjacencies}, for which we give several constant ratio
  approximation algorithms in the maximum matching model, in the case where genomes contain
  the same number of duplications of each gene.

  \vspace{\baselineskip}

\end{abstract}

\section{Introduction and Preliminaries}

In comparative genomics, computing a measure of (dis-)similarity
between two genomes is a central problem: such a measure can be used,
for instance, to construct phylogenetic trees. The measures defined so
far essentially fall into two categories: the first one consists in counting the
minimum number of operations needed to transform a genome 
into another (e.g. the \emph{edit distance}~\cite{Swenson_LCS}
or the \emph{number of reversals}~\cite{reversal_distance}). The second
one contains (dis-)similarity measures based on the genome structure,
such as the 
\emph{number of breakpoints}~\cite{Complexite_exemplarisation}, 
the \emph{conserved intervals distance}~\cite{NP_Completude_E_MConsI}, 
the \emph{number of common
  intervals}~\cite{NP_completude_EComI_MComI}, 
\emph{SAD} and
\emph{MAD}~\cite{Introduction_MAD_SAD} etc.  

When genomes contain no duplicates, most measures can be computed in polynomial
time. However, assuming that genomes contain no duplicates is too
limited. Indeed, it has been recently shown that a great number of
duplicates exists in some genomes. For 
example, in~\cite{duplications_gene_humain}, authors estimate that
$15\%$ of genes are duplicated in the human genome. A possible approach to 
overcome this difficulty is to specify a one-to-one correspondence (i.e. a
{\em matching}) between genes of both genomes and to remove the
unmatched genes, thus obtaining two genomes with identical gene
content and no duplicates. Usually, the above mentioned matching is
chosen in order to optimize the studied measure, following the
parsimony principle. Three models   
achieving this correspondence have been proposed : the \emph{exemplar}
model~\cite{Sankoff_rearrangement}, the \emph{intermediate} 
model~\cite{article_JCB_pseudo_booleen_IC} and the \emph{maximum matching}
model~\cite{introduction_matching}. Before defining precisely the
measures and models studied in this paper, we need to introduce some notations.

\paragraph{Notations used in the paper.}


A genome $G$ is represented by a sequence of signed
integers (called \emph{signed genes}). For any
genome $G$, we denote by $\mathcal{F}_G$ the set of unsigned integers
(called \emph{genes}) that are present in $G$.
For any signed gene $g$, let $-g$ be the 
signed gene having the opposite sign and let $|g|\in \mathcal{F}_G$ be the
corresponding (unsigned) gene.

Given a genome $G$ without duplicates 
and two signed genes $a$, $b$ such that 
$a$ is located before $b$, let $G[a,b]$ be the set $S\subseteq
\mathcal{F}_G$ of genes located between 
genes $a$ and $b$ in $G$, $a$ and $b$ included. We also note $[a,b]_G$ the substring (i.e. the
sequence of consecutive elements) of $G$ starting at $a$ and finishing
at $b$ in $G$.

Let $\OCC(g,G)$ be the number of occurrences of a
given gene $g$ in a genome $G$ and let $\OCC(G)= \max\{\OCC(g,G)| g\in \mathcal{F}_G\}$. 
A pair of genomes ($G_1$,$G_2$) is said to be {\em of type $(x,y)$} if
$\OCC(G_1)=x$ and $\OCC(G_2)=y$. A pair of genomes ($G_1$,$G_2$) is said to be \emph{balanced} if, for
each gene $g\in \mathcal{F}_{G_1}\cup \mathcal{F}_{G_2}$, we have
$\OCC(g,G_1) = \OCC(g,G_2)$ (otherwise, ($G_1$,$G_2$) will be said to be
{\em unbalanced}). Note that a pair $(G_1,G_2)$ of type $(x,x)$ is
not necessary balanced. 

Denote by $n_G$ the size of genome $G$, that is the number of signed genes it contains.  
Let $G[p]$, $1 \leq p \leq n_G$, be the signed gene that occurs at
position $p$ on genome $G$, and let $\vert G[p]\vert\in \mathcal{F}_G$ be the
corresponding (unsigned) gene. Let $N_G[p]$, $1 \leq p \leq n_G$, be 
the number of occurrences of $\vert G[p]\vert$ in the first $(p-1)$
positions of $G$. 

We define a 
\emph{duo} in a genome $G$ as a pair of successive signed genes.
Given a duo $d_i = (G[i],G[i+1])$ in a
genome $G$, we note $-d_i$ the duo equal  
to $(-G[i+1],-G[i])$. Let $(d_1,d_2)$ be a pair
of duos~; $(d_1,d_2)$ is called a \emph{duo match} if $d_1$ is a
duo of $G_1$, $d_2$ is a duo of $G_2$, and if either $d_1=d_2$ or
$d_1=-d_2$.

For example, consider the genome $G_1 = +1~+2~+3~+4~+5~-1~-2~+6~-2$.
Then, $\mathcal{F}_G =
\{1,2,3,4,5,6\}$, $n_{G_1}=9$, $\OCC(1,G_1)=2$, $\OCC(G_1)=3$, $G_1[7]=-2$,  
$-G_1[7]=+2$, $|G_1[7]|=2$ and $N_{G_1}[7]=1$. Let $G_2$ be the genome $G_2 = +2~-1~+6~+3~-5~-4~+2~-1~-2$. 
Then the pair ($G_1$, $G_2$) is balanced and is of type $(3,3)$.
Let $d_1=(G_1[4],G_1[5])$ be the duo $(+4,+5)$ and $d_2$ be the duo $(G_2[5],G_2[6])$.
The pair $(d_1,d_2)$ is a duo match. Now, consider the genome $G_3 = +3 -2 +6 +4 -1 +5$ without
duplicates. We have $G_3[+6,-1]= \{1,4,6\}$ and
$[+6,-1]_{G_3}=(+6,+4,-1)$.

\paragraph{Breakpoints, adjacencies, common and conserved intervals.}

Let us now define the four measures we will study in this paper. Let
$G_1$ and $G_2$ be two genomes without duplicates and with the same gene
content, that is $\mathcal{F}_{G_1}=\mathcal{F}_{G_2}$. 

\emph{Breakpoint and Adjacency.} 
Let $(a,b)$ be a duo in $G_1$. We say 
that the duo $(a, b)$ induces a \emph{breakpoint} of $(G_1,G_2)$ if 
neither $(a,b)$ nor $(-b,-a)$ is a duo in
$G_2$. Otherwise, we say that $(a,b)$ induces an \emph{adjacency} of
$(G_1,G_2)$. For example, when $G_1 = +1 +2 +3 +4 +5$ and $G_2 = +5 -4
-3 +2 +1$, the duo $(2,3)$ in $G_1$ induces a breakpoint of
$(G_1,G_2)$ while $(3,4)$ in $G_1$ induces an adjacency of
$(G_1,G_2)$. We note $B(G_1,G_2)$ (resp. $A(G_1,G_2)$) the number of
breakpoints (resp. the number of adjacencies) that exist
between $G_1$ and $G_2$.


\emph{Common interval.} A \emph{common interval} of $(G_1,G_2)$ is a
substring of $G_1$ such that $G_2$ contains a permutation of this
substring (not taking signs into account). For example, consider $G_1
= +1 +2 +3 +4 +5$ and $G_2 = +2 -4 +3 +5 +1$. The substring
$[+3,+5]_{G_1}$ is a common  interval of $(G_1,G_2)$. 

\emph{Conserved interval.} 
Consider two signed genes $a$ and $b$ of
$G_1$ such that $a$ precedes $b$, where the precedence relation is
large in the sense that, possibly, $a=b$. The substring $[a,b]_{G_1}$
is a \emph{conserved interval} of $(G_1,G_2)$ if either (i)~$a$
precedes $b$ and $G_2[a,b]=G_1[a,b]$, or~(ii)
$-b$ precedes $-a$ and $G_2[-b,-a]=G_1[a,b]$. 
For example, if $G_1 = +1 +2 +3 +4 +5$ 
and $G_2 = -5 -4 +3 -2 +1$, the substring $[+2,+5]_{G_1}$ is a 
conserved interval of $(G_1,G_2)$. We note that the notion of 
conserved interval does not consider the sign of genes. Note also that a
conserved interval is actually a common interval, but with additional
restrictions on its extremities.


\paragraph{Dealing with duplicates in genomes.}

When genomes contain duplicates, we cannot directly compute the
measures defined in the previous paragraph. A solution consists in finding a
one-to-one correspondence  (i.e. a matching) between duplicated genes
of $G_1$ and $G_2$~; we then use this correspondence to rename genes of
$G_1$ and $G_2$, and we delete the unmatched signed genes in order to
obtain two genomes $G'_1$ and $G'_2$ such that $G'_2$ is a
{\em permutation} of $G'_1$~; thus, the measure computation becomes
possible. In this paper, we will focus on three models of matching :
the \emph{exemplar}, \emph{intermediate} and \emph{maximum matching}
models.

\begin{itemize} 
\item The \emph{exemplar model}~\cite{Sankoff_rearrangement}: for each
  gene $g$, we keep in the matching $\mathcal{M}$ only one occurrence of $g$ in
  $G_1$ and in $G_2$, and we remove all the other occurrences. Hence,
  we obtain two genomes $G_1^E$ and $G_2^E$ without duplicates. The
  triplet $(G_1^E,G_2^E,\mathcal{M})$  is called an \emph{exemplarization} of
  $(G_1,G_2)$. Note that in this model, $\mathcal{M}$ can
  be inferred
  from the exemplarized genomes $G_1^E$ and $G_2^E$. Thus, in the rest
  of the paper, any exemplarization $(G_1^E,G_2^E,\mathcal{M})$ of $(G_1,G_2)$
  will be only described by the pair $(G_1^E,G_2^E)$.

\item The \emph{intermediate model}~\cite{article_JCB_pseudo_booleen_IC}: in this model, for each gene
  $g$, we keep in the matching $\mathcal{M}$ an arbitrary number $k_g$, $1 \leq k_g
  \leq min(occ(g,G_1),occ(g,G_2))$, in order to obtain genomes $G_1^I$ and
  $G_2^I$. We call the triplet $(G_1^I,G_2^I,\mathcal{M})$ an \emph{intermediate
    matching} of $(G_1,G_2)$.

\item The \emph{maximum matching model}~\cite{introduction_matching}:
  in this case, we keep in the matching $\mathcal{M}$ the maximum number of signed genes in
  both genomes. More precisely, we look for a one-to-one
  correspondence between signed genes of $G_1$ and $G_2$ that matches, for each gene
  $g$, exactly $min(\OCC(g,G_1), $ \linebreak $\OCC(g,G_2))$ occurrences. After
  this operation, we delete each unmatched signed gene. The triplet
  $(G_1^M,G_2^M,\mathcal{M})$ obtained by this operation is called a \emph{maximum
    matching} of $(G_1,G_2)$.

\end{itemize}


\paragraph{Problems studied in this paper.}


Consider two genomes $G_1$ and $G_2$ with duplicates. 
Let \PB{EComI} (resp. \PB{IComI}, \PB{MComI}) be the problem which consists in
finding an exemplarization (resp. intermediate matching, maximum matching) $(G'_1,G'_2,\mathcal{M})$ of
$(G_1,G_2)$ such that the number of common intervals of $(G'_1,G'_2)$
is maximized. Moreover, let \PB{EConsI} (resp. \PB{IConsI}, \PB{MConsI}) be the problem which consists
in finding an exemplarization 
(resp. intermediate matching, maximum matching) $(G'_1,G'_2,\mathcal{M})$ of
$(G_1,G_2)$ such that the number of
conserved intervals of $(G'_1,G'_2,\mathcal{M})$ is maximized.  
In Section \ref{Hardness_intervals}, we prove the \APXhardness of \PB{EComI} 
and \PB{EConsI}, even for genomes $G_1$ and $G_2$ such that 
$\OCC(G_1)=1$ and $\OCC(G_2)=2$. These results induce the \APXhardness under the
other models (i.e., \PB{IComI}, \PB{MComI}, \PB{IConsI} and
\PB{MConsI} are \APXhard). These results extend in particular those
of~\cite{Complexite_exemplarisation,NP_completude_EComI_MComI}.

Let \PB{EBD} (resp. \PB{IBD}, \PB{MBD}) be the problem which consists
in finding  an exemplarization 
(resp. intermediate matching, maximum
matching) $(G'_1,G'_2,\mathcal{M})$ of $(G_1,G_2)$ that minimizes the number of
breakpoints between $G'_1$ and 
$G'_2$. In Section~\ref{Hardness_breakpoints}, we prove the \APXhardness
of \PB{EBD}, even for genomes $G_1$ and $G_2$ such that  
$\OCC(G_1)=1$ and $\OCC(G_2)=2$. This result implies that \PB{IBD} and
\PB{MBD} are also \APXhard, and extends those
of~\cite{chen_inproceding_approximation_breakpoint}. 

Let \PB{ZEBD} (resp. \PB{ZIBD}, \PB{ZMBD}) be the problem which
consists in determining, for two genomes $G_1$ and $G_2$, whether there exists an 
exemplarization (resp. intermediate matching, maximum matching) which
induces {\em zero breakpoint}. In section~\ref{Complexity_ZXBD}, we
study the complexity of \PB{ZEBD},  
\PB{ZMBD} and \PB{ZIBD}: in particular, we extend a result
of~\cite{chen_inproceding_approximation_breakpoint} by proving
\PB{ZEBD} to be \NPC for a new class of instances. We also note that
the problems \PB{ZEBD} and \PB{ZIBD} are equivalent, and we show that \PB{ZMBD} is in
\Pclass.

Finally, in Section \ref{Approximation_adjacences}, we focus on a
fourth measure, closely related to the number of breakpoints: the
\emph{number of  adjacencies}, for which we give several constant ratio
approximation algorithms in the maximum matching model, in the case
where genomes are balanced.

\section{\PB{EComI} and \PB{EConsI} are \APXhard}
\label{Hardness_intervals}

Consider two genomes $G_1$ and $G_2$ with duplicates, and let
\PB{EComI} (resp. \PB{IComI}, \PB{MComI}) be the problem which
consists in finding an exemplarization (resp. intermediate matching, maximum matching) $(G'_1,G'_2,\mathcal{M})$ of
$(G_1,G_2)$ such that the number of common intervals of $(G'_1,G'_2)$
is maximized. Moreover, let \PB{EConsI} (resp. \PB{IConsI}, \PB{MConsI}) be the problem which consists
in finding an exemplarization 
(resp. intermediate matching, maximum matching) $(G'_1,G'_2,\mathcal{M})$ of
$(G_1,G_2)$ such that the number of
conserved intervals of $(G'_1,G'_2,\mathcal{M})$ is maximized.

\PB{EComI} and \PB{MComI} have been proved to be \NPC even if $\OCC(G_1) = 1$ and 
$\OCC(G_2) = 2$ in~\cite{NP_completude_EComI_MComI}.  
Besides, in~\cite{NP_Completude_E_MConsI}, Blin and Rizzi have studied the problem of
computing a distance built on the number of conserved intervals. This
distance differs from the number of conserved intervals we study in this 
paper, mainly in the sense that (i)~it can be applied to two {\em sets} 
of genomes (as opposed to two genomes in our case), and (ii)~the 
distance between two identical genomes of length $n$ is equal to 0 (as 
opposed to $\frac{n(n+1)}{2}$ in our case). Blin and 
Rizzi~\cite{NP_Completude_E_MConsI} proved that finding the minimum distance is 
\NPC, under both the exemplar and maximum matching
models. A closer analysis of their proof shows that it can be easily adapted to 
prove that \PB{EConsI} and \PB{MConsI} are \NPC, even in the case 
$\OCC(G_1)=1$.

We can conclude from the above results that \PB{IComI} and \PB{IConsI} are also \NPC, since when one 
genome contains no duplicates, \emph{exemplar}, \emph{intermediate}
and \emph{maximum matching} models are equivalent.

In this section, we improve the above results by showing that the six
problems \PB{EComI}, \PB{IComI}, \PB{MComI}, \PB{EConsI}, \PB{IConsI}
and \PB{MConsI} are \APXhard, even when genomes $G_1$ and $G_2$ are
such that $\OCC(G_1) = 1$ and $\OCC(G_2) = 2$. The main 
result is Theorem~\ref{theoreme1}, which will be completed by Corollary~\ref{corollary_interval}
at the end of the section.

\begin{theorem} \label{theoreme1}
  \PB{EComI} and \PB{EConsI} are \APXhard even when 
  genomes $G_1$ and $G_2$ are such that $\OCC(G_1) = 1$ and $\OCC(G_2) = 2$.  
\end{theorem}

We prove Theorem~\ref{theoreme1} by using an
\emph{L-reduction}~\cite{APX_completude_papadimitriou} from the
\PB{Min-Vertex-Cover} problem on cubic graphs, denoted here
\PB{Min-Vertex-Cover-3}. Let $G=(V,E)$ be a cubic graph, i.e. for all $v \in V,
degree(v)=3$. A set of vertices $V' \subseteq V$ is called a
\emph{vertex cover} of $G$ if for each edge $e \in E$, there exists a
vertex $v \in V'$ such that $e$ is incident to $v$. The problem \PB{Min-Vertex-Cover-3}
is defined as follows:

\medskip
\mabox{\textbf{Problem:} \PB{Min-Vertex-Cover-3}\\
  \textbf{Input:} A cubic graph $G=(V,E)$.\\
  \textbf{Solution:} A vertex cover $V'$ of $G$. \\
  \textbf{Measure:} The cardinality of $V'$.
}
\medskip

\PB{Min-Vertex-Cover-3} was proved to be \APXC
in~\cite{APX_VERTEX_COVER_CUBIC}. 

\subsection{Reduction}

Let $G=(V,E)$ be an instance of \PB{Min-Vertex-Cover-3}, where $G$ is a cubic graph
with $V = \{v_1\ldots v_n\}$ and $E = \{e_1\ldots e_m\}$. Consider the
transformation $R$ which associates to the graph $G$ two genomes $G_1$
and $G_2$ in the following way, where each gene has a positive sign. 
\begin{align}
  G_1 &=  b_1 ~ b_2 \ldots b_m ~ x ~ a_1 ~C_1 ~f_1 ~ a_2 ~ C_2 ~ f_2
  \ldots a_n ~ C_n ~f_n ~ y ~ b_{m+n},b_{m+n-1} \ldots b_{m+1} \\
  G_2 &= y ~ a_1 ~ D_1 ~ f_1 ~ b_{m+1} ~ a_2 ~ D_2 ~ f_2 ~ b_{m+2}
  \ldots b_{m+n-1} ~ a_n ~ D_n ~ f_n ~ b_{m+n} ~x
\end{align}
with : 
\begin{itemize}
\item for each $i$, $1 \leq i \leq n, a_i = 6i - 5$, $f_i = 6i$
\item  for each $i$, $1 \leq i \leq n, C_i = (a_i + 1), (a_i + 2),(a_i + 3), (a_i + 4)$
\item for each $i$, $1 \leq i \leq n+m, b_i = 6n +i$
\item $x = 7n + m + 1$ and $y = 7n + m + 2$
\item for each $i$, $1 \leq i \leq n, D_i = (a_i + 3), (b_{j_i}), (a_i + 1), (b_{k_i}), (a_i + 4), (b_{l_i}), (a_i + 2)$ where $e_{j_i}$, $e_{k_i}$ and $e_{l_i}$ are the edges which are incident to $v_i$ in $G$, with $j_i < k_i < l_i$.
\end{itemize}
In the following, genes $b_i$, $1 \leq i \leq m$, are called \emph{markers}. There is no duplicated gene in $G_1$ and the markers are the only duplicated genes in $G_2$~; these genes occur twice  in $G_2$. Hence, we have $\OCC(G_1) = 1$ and $\OCC(G_2) = 2$. 


\begin{figure}
  \begin{center}
    \input{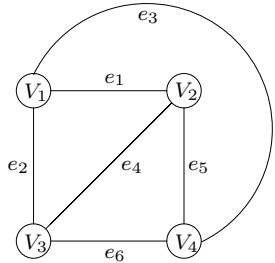}
  \end{center}
  \caption{\label{cubic_graph} The cubic graph $G$.} 
\end{figure}

To illustrate the reduction, consider the cubic graph $G$ of Figure \ref{cubic_graph}. From $G$, we construct the following genomes $G_1$ and $G_2$: 
\scriptsize{
  \begin{center}
    $\overbrace{25}^{b_1}\overbrace{26}^{b_2}\overbrace{27}^{b_3}\overbrace{28}^{b_4}\overbrace{29}^{b_5}\overbrace{30}^{b_6} \overbrace{35}^{x}  1 \overbrace{2~3~4~5}^{C_1} 6 ~ 7 \overbrace{8~9~10~11}^{C_2} ~ 12 ~ 13 ~ \overbrace{14~15~16~17 }^{C_3} ~18~19~ \overbrace{20~21~22~23}^{C_4} ~24 \overbrace{36}^{y} \overbrace{34}^{b_{10}}\overbrace{33}^{b_9}\overbrace{32}^{b_8}\overbrace{31}^{b_7}$ \\ 
    ~\\
    $\underbrace{36}_y  1 \underbrace{4~25~2~26~5~27~3}_{D_1} 6 \underbrace{31}_{b_7} 7 \underbrace{10~25~8~28~11~29~9}_{D_2} 12 \underbrace{32}_{b_8}  13 \underbrace{16~26~14~28~17~30~15}_{D_3} 18  \underbrace{33}_{b_9}  19 \underbrace{22~27~20~29~23~30~21}_{D_4} 24  \underbrace{34}_{b_{10}} \underbrace{35}_x$
  \end{center}}
\normalsize

\subsection{Preliminary results}
\label{Preliminary_results_Intervals}

In order to prove Theorem \ref{theoreme1}, we first give four intermediate lemmas. 
In the following, a common interval for the \PB{EComI} problem or a
conserved interval for \PB{EConsI} is called a \emph{robust
  interval}. Besides, a \emph{trivial interval} will denote either an
interval of length one (i.e. a singleton), or the whole genome.

\begin{lemma} \label{lemme1}
  For any exemplarization $(G_1,G_2^E)$  of $(G_1,G_2)$, the non trivial robust intervals of \linebreak 
  $(G_1,G_2^E)$ are necessarily contained in some sequence $a_i C_i f_i$ of $G_1$ ($1 \leq i \leq n$).
\end{lemma}

\begin{proof}
  We start by proving the lemma for common intervals, and we will then extend it to conserved intervals.
  First, we prove that, for any exemplarization $(G_1,G_2^E)$ of $(G_1,G_2)$, each common interval $I$ such 
  that $|I| \geq 2$ contains either both of $x$, $y$ or none of them.  This
  further  implies that $I$ covers the whole genome.
  Suppose there exists a common interval $I_x$ (recall that by definition $I_x$ is on $G_1$) 
  such that $|I_x| \geq 2$ and $I_x$ contains $x$. Let $PI_x$ be the permutation of $I_x$ in $G_2^E$. The interval $I_x$ must contain either $b_m$ or $a_1$. Let us detail each of the two cases:
  \begin{itemize}
  \item[(a)] If $I_x$ contains $b_m$, then $PI_x$ contains $b_m$ too. Notice that there is some $i$, $1 \leq i \leq n$, such that $b_m$  belongs to $D_i$ in $G_2^E$. Then $PI_x$ contains all genes between $D_i$ and $x$ in $G_2^E$. Thus $PI_x$ contains $b_{m+n}$. Consequently, $I_x$ contains $b_{m+n}$ and it also contains $y$.
  \item[(b)] If $I_x$ contains $a_1$, then $PI_x$ contains $a_1$ too. Then $PI_x$ contains all genes between $a_1$ and $x$. Thus $PI_x$ contains $b_{m+n}$. Hence, $I_x$ contains $b_{m+n}$ and then it also contains $y$.
  \end{itemize}

  Now, suppose that $I_y$ is a common interval such that $|I_y| \geq 2$ and $I_y$ contains $y$. Let $PI_y$ be the permutation of $I_y$ on $G_2^E$. The interval $I_y$ must contain either $b_{m+n}$ or $f_n$. Let us detail each of the two cases:
  \begin{itemize}
  \item[(a)] If $I_y$ contains $b_{m+n}$, then $PI_y$ contains $b_{m+n}$ too. Thus $PI_y$ contains all genes between $b_{m+n}$ and $y$. Hence $PI_y$ contains all the sequences $D_i$, $1 \leq i \leq n$. In particular, $PI_y$ contains all the markers and consequently $I_y$ must contain $x$.
  \item[(b)] If $I_y$ contains $f_n$, then $PI_y$ contains $f_n$ too. Then $PI_y$ contains all genes between $f_n$ and $y$. In particular, $PI_y$  contains $b_{m+n-1}$  and then $I_y$ contains $b_{m+n-1}$ too. Hence, $I_y$ also contains $b_{m+n}$, similarly to the previous case. Thus $I_y$ contains $x$.
  \end{itemize}

  We conclude that each non singleton common interval containing either $x$ or $y$ necessarily contains both $x$ and $y$. Therefore, and by construction of $G_2$, there is only one such interval, that is $G_1$ itself. Hence, any non trivial common interval is necessarily, in $G_1$, either strictly on the left of $x$, or between $x$ and $y$, or strictly on the right of $y$. Let us analyze these different cases:

  \begin{itemize}
  \item Let $I$ be a non trivial common interval situated strictly on the left of $x$ in $G_1$.	Thus $I$ is a sequence of at least two consecutive markers. Since in any exemplarization $(G_1,G_2^E)$ of $(G_1,G_2)$, every marker in $G_2^E$ has neighboring genes which are not markers, this contradicts the fact that $I$ is a common interval.
  \item Let $I$ be a non trivial common interval situated strictly on the right of $y$ in $G_1$. Then $I$ is a substring of $b_{m+n}, \ldots, b_{m+1}$ containing at least two genes. In any exemplarization $(G_1,G_2^E)$ of $(G_1,G_2)$, for each pair $(b_{m+i},b_{m+i+1})$ of $G_2^E$, with $1 \leq i < n$, we have $a_{i+1} \in G_2^E[b_{m+i},b_{m+i+1}]$. This contradicts the fact that $I$ is strictly on the right of $y$ in $G_1$.
  \item  Let $I$ be a non trivial common interval lying between $x$ and $y$ in $G_1$.	For any exemplarization $(G_1,G_2^E)$ of $(G_1,G_2)$, a common interval cannot contain, in $G_1$, both $f_i$ and $a_{i+1}$ for some $i$, $1 \leq i \leq n-1$ (since $b_{m+i}$ is situated between $f_i$ and $a_{i+1}$ in $G_2^E$ and on the right of $x$ in $G_1$). Hence, a non trivial common interval of $(G_1,G_2^E)$ is included in some sequence $a_iC_if_i$ in $G_1$, $1 \leq i \leq n$.
  \end{itemize}  

  This proves the lemma for common intervals.  By definition, any conserved interval is necessarily a common
  interval. So, a non trivial conserved interval of $(G_1,G_2^E)$ is
  included in some sequence $a_iC_if_i$ in $G_1$, $1 \leq i
  \leq n$. 
  The lemma is proved. \qed  

\end{proof}

\begin{lemma} \label{lemme2}
  Let $(G_1,G_2^E)$ be an exemplarization of $(G_1,G_2)$ and $i \in [1
  \ldots n]$. Let $\Delta_i$ be a substring of $[a_i+3,a_i+2]_{G_2^E}$
  that does not contain any marker. If $| \Delta_i | \in \{2, 3\}$, then
  there is no robust interval $I$ of $(G_1,G_2^E)$ such that $\Delta_i$
  is a permutation of $I$.  
\end{lemma}

\begin{proof}
  First, we prove that there is no permutation $I$ of $\Delta_i$ such
  that $I$ is a common interval of $(G_1,G_2^E)$. Next, we show that
  there is no permutation $I$ of $\Delta_i$ such that $I$ is a conserved
  interval. By Lemma \ref{lemme1}, we know that a non trivial common
  interval of $(G_1,G_2^E)$ is a substring of some sequence $a_iC_if_i$,
  $1 \leq i \leq n$. This substring contains only consecutive
  integers. Therefore, if there exists a permutation $I$ of $\Delta_i$
  such that $I$ is a common interval of $(G_1,G_2^E)$, then $\Delta_i$
  must be a permutation of consecutive integers. If $| \Delta_i | = 2$,
  we have $\Delta_i = (p,q)$ where $p$ and $q$ are not consecutive
  integers and if $| \Delta_i | = 3$, then we have $\Delta_i =
  (a_i+3,a_i+1,a_i+4)$  or $\Delta_i = (a_i+1,a_i+4,a_i+2)$. In these
  three cases, $\Delta_i$ is not a permutation of consecutive
  integers. Hence, there is no permutation $I$ of $\Delta_i$ such that
  $I$ is a common interval of $(G_1,G_2^E)$. Moreover, any conserved
  interval is also a common interval. Thus, there is no permutation $I$
  of $\Delta_i$ such that $I$ is a conserved interval of
  $(G_1,G_2^E)$. \qed 
\end{proof}

For more clarity, let us now introduce some notations. Given a graph
$G = (V,E)$, let $VC = \{v_{i_1},v_{i_2}\ldots v_{i_k}\}$ be a  vertex
cover of $G$. Let $R(G) = (G_1,G_2)$ be the pair of genomes defined by
the construction described in $(1)$ and $(2)$. Now, let $F$ be the
function which associates to $VC$, $G_1$ and $G_2$ an exemplarization
$F(VC)$ of $(G_1,G_2)$ as follows. In $G_2$, all the markers are
removed from the sequences $D_i$ for all $i \neq i_1,i_2 \ldots
i_k$. Next, for each marker which is still present twice, one of its
occurrences is arbitrarily removed. Since in $G_2$ only markers are
duplicated, we conclude that $F(VC)$ is an exemplarization of
$(G_1,G_2)$.  

Given a cubic graph $G$ and genomes $G_1$ and $G_2$ obtained by the
transformation $R(G)$, let us define the function $S$ which associates
to an exemplarization $(G_1,G_2^E)$ of $(G_1,G_2)$ the vertex cover
$VC$ of $G$ defined as follows: $VC = \{v_i| 1\leq i \leq
n \wedge \exists j \in \{1 \ldots m\}, b_j \in G_2^E[a_i,f_i]\}$. In
other words, we keep in $VC$ the vertices $v_i$ of $G$ for which
there exists some gene $b_j$ such that $b_j$ is in
$G_2^E[a_i,f_i]$. We now prove that $VC$ is a vertex
cover. Consider an edge $e_p$ of $G$. By construction of $G_1$ and
$G_2$, there exists some $i$, $1 \leq i \leq n$, such that
gene $b_p$ is located between $a_i$ and $f_i$ in $G_2^E$. The presence
of gene $b_p$ between  $a_i$ and $f_i$ implies that vertex $v_i$
belongs to $VC$. We conclude that each edge is incident to at least
one vertex of $VC$.  

Let $W$ be the function defined on $\{\PB{EConsI}, \PB{EComI}\}$ by $W(\text{pb})=1$ if
$\text{pb} = \PB{EConsI}$ and $W(\text{pb})=4$ if $\text{pb} = \PB{EComI}$. Let $\OPT_P(A)$ be the optimum
result of an instance $A$ for an optimization problem $\text{pb}$, 
$\text{pb}\in \{\PB{EcomI}, \PB{EConsI}, \PB{Min-Vertex-Cover-3}\}$. 

We now define the function $T$ whose arguments are a problem $\text{pb} \in
\{\PB{EConsI}, \PB{EComI}\}$ and a cubic graph $G $. Let $R(G)=(G_1, G_2^E)$
as usual. Then $T(\text{pb}, G)$ is defined as the number of
robust trivial intervals of $(G_1, G_2^E)$ with respect to $\text{pb}$. Let $n$ and $m$ be
respectively the number of vertices and the number of edges of $G$. We
have $T(\PB{EConsI},G) = 7n + m + 2$ and $T(\PB{EComI},G) = 7n + m +
3$. Indeed, for \PB{EComI}, there are $7n+m+2$ singletons and we also
need to consider the whole genome. 

\begin{lemma} \label{lemme3}
  Let $\text{pb} \in \{\PB{EcomI}, \PB{EConsI}\}$. Let $G$ be a cubic graph and
  $R(G) = (G_1,G_2)$. Let $(G_1,G_2^E)$ be an exemplarization of
  $(G_1,G_2)$ and let $i$, $1 \leq i \leq n$. Then only two
  cases can occur with respect to $D_i$. 
  \begin{enumerate} 
  \item Either in $G_2^E$, all the markers from $D_i$ were
    removed, and in this case, there are exactly $W(\text{pb})$ non
    trivial robust intervals involving $D_i$. 
  \item  Or in $G_2^E$, at least one marker was kept in $D_i$,
    and in this case, there is no non trivial robust interval
    involving $D_i$. 
  \end{enumerate} 
\end{lemma}

\begin{proof}
  We first prove the lemma for the \PB{EComI} problem and then we extend
  it to \PB{EConsI}. Lemma~\ref{lemme1} implies that each non trivial
  common interval $I$ of $(G_1,G_2^E)$ is contained in some substring of
  $a_iC_if_i$, $1 \leq i \leq n$. So, the permutation of $I$
  on $G_2^E$ is contained in a substring of $a_i D_if_i$, $1 \leq i
  \leq n$. 

  Consider $i$, $1 \leq i \leq n$, and suppose that
  all the markers from $D_i$ are removed on $G_2^E$. Thus, $a_iC_if_i$,
  $C_i$, $a_iC_i$ and $C_if_i$ are common intervals of
  $(G_1,G_2^E)$. Let us now show that there is no other non trivial
  common interval involving  $D_i$. Let $\Delta_i$ be a substring of
  $[a_i+3,a_i+2]_{G_2^E}$ such that $| \Delta_i | \in \{2,3\}$. By Lemma
  \ref{lemme2}, we know that $\Delta_i$ is not a common interval. The
  remaining intervals are $(a_i,a_i+3)$, $(a_i,a_i+3,a_i+1)$,
  $(a_i,a_i+3,a_i+1,a_i+4)$, $(a_i+1,a_i+4,a_i+2,f_i)$,
  $(a_i+4,a_i+2,f_i)$ and $(a_i+2,f_i)$. 
  By construction, none of them can be a common interval, because none
  of them is a permutation of consecutive integers. Hence, there are
  only four non trivial common intervals involving $D_i$ in
  $G_2^E$. Among these four common intervals, only $a_iC_if_i$ is a
  conserved interval too. In the end, if all the markers are removed
  from $D_i$, there are exactly four non trivial common intervals and
  one non trivial conserved interval involving  $D_i$. So, given a
  problem $\text{pb} \in \{\PB{EcomI}, \PB{EconsI}\}$, there are exactly $W(\text{pb})$ non
  trivial robust intervals involving  $D_i$. 

  Now, suppose that at least one marker of $D_i$ is kept in
  $G_2^E$. Lemma \ref{lemme1} shows that each non trivial common
  interval $I$ of $(G_1,G_2^E)$ is contained in some substring of
  $a_iC_if_i$, $1 \leq i \leq n$. Since no marker is present
  in a sequence $a_iC_if_i$, we deduce that there does not exist any
  trivial common interval containing a marker. So, a non trivial common
  interval involving $D_i$ only must contain a substring  $\Delta_i$ of
  $[a_i+3,a_i+2]_{G_2^E}$ such that $\Delta_i$ contains no marker. Since
  no marker is an extremity of $[a_i+3,a_i+2]_{G_2^E}$, we have $|
  \Delta_i | \leq 3$. By Lemma  \ref{lemme2}, we know that
  $\Delta_i$ is not a common interval. The remaining intervals to be
  considered  are the intervals $a_i\Delta_i$ and $\Delta_if_i$. By
  construction of  $a_iC_if_i$, these intervals are not common intervals
  (the absence of gene $a_i+2$ for $a_i\Delta_i$ and of gene $a_i+3$ for
  $\Delta_if_i$ implies that these intervals are not a permutation of
  consecutive integers). Hence, these intervals cannot be conserved
  intervals either. \qed 
\end{proof}

\begin{lemma} \label{lemme4}
  Let $\text{pb} \in \{\PB{EcomI}, \PB{EConsI}\}$. Let $G = (V,E)$ be a cubic
  graph with $V=\{v_1\ldots v_n\}$ and $E=\{e_1\ldots e_m\}$ and let
  $G_1$, $G_2$ be the two genomes obtained by $R(G)$. 
  \begin{enumerate}
  \item Let $VC$ be a vertex cover of $G$ and denote $k = |VC|$. Then
    the exemplarization $F(VC)$ of $(G_1,G_2)$ has at least $N =
    n\,W(\text{pb}) + T(\text{pb},G) - W(\text{pb})\cdot k$ robust intervals. 
  \item Let $(G_1,G_2^E)$ be an exemplarization of $(G_1,G_2)$ and let
    $VC'$ be the vertex cover of $G$ obtained by $S(G_1,G_2^E)$. Then
    $|VC'|= \frac{W(\text{pb})\cdot n + T(\text{pb},G) - N}{W(\text{pb})}$, where $N$ is the
    number of robust intervals of $(G_1,G_2^E)$. 
  \end{enumerate}
\end{lemma}

\begin{proof}
  1. Let $\text{pb} \in \{\PB{EcomI}, \PB{EConsI}\}$. Let $G$ be a cubic graph and let
  $G_1$ and $G_2$ be the two genomes obtained by $R(G)$. Let
  $VC$ be a vertex cover of $G$ and denote  $k = |VC|$. Let
  $(G_1,G_2^E)$ be the exemplarization of $(G_1,G_2)$ obtained by
  $F(VC)$. By construction, we have at least $(n-k)$ substrings $D_i$ in
  $G_2^E$ for which all the markers are removed. By Lemma  \ref{lemme3},
  we know that each of these substrings implies the existence of $W(\text{pb})$
  non trivial robust intervals. So, we have at least $W(\text{pb})(n-k)$ non
  trivial robust intervals. Moreover, by definition of $T(\text{pb}, G)$, the number
  of trivial robust intervals of $(G_1,G_2^E)$ is exactly
  $T(\text{pb},G)$. Thus, we have at least $N = W(\text{pb})\cdot n + T(\text{pb},G)
  -W(\text{pb})\cdot k$ robust intervals of $(G_1,G_2^E)$.  

  2. Let $(G_1,G_2^E)$ be an exemplarization of $(G_1,G_2)$ and let $n-j$ be
  the number of sequences $D_i$, $1 \leq i \leq n$, for
  which all markers have been deleted in $G_2^E$. Then, by
  Lemmas~\ref{lemme1} and~\ref{lemme3}, the number of robust
  intervals of $(G_1,G_2^E)$ is equal to $N = W(\text{pb})\cdot n + T(\text{pb},G)
  - W(\text{pb})\cdot j$. Let $VC'$ be the vertex cover obtained by
  $S(G_1,G_2^E)$. Each marker has one occurrence in $G_2^E$ and these
  occurrences lie in $j$ sequences $D_i$. So, by definition of $S$,
  we conclude that $|VC'|= j = \frac{W(\text{pb})\cdot n + T(\text{pb},G) -
    N}{W(\text{pb})}$. \qed 
\end{proof}

\subsection{Main result} 
Let us first define the notion of
\emph{L-reduction}~\cite{APX_completude_papadimitriou}: let $A$ and
$B$ be two optimization problems and $c_A$, $c_B$ be respectively
their cost functions. An \emph{L-reduction} from problem $A$ to
problem $B$ is a pair of polynomial-time computable functions $R$ and $S$ with the
following properties: 
\begin{itemize}
\item[$(a)$] If $x$ is an instance of $A$, then $R(x)$ is an instance of $B$~;
\item[$(b)$] If $x$ is an instance of $A$ and $y$ is a solution of $R(x)$, then $S(y)$ is a solution of  $A$~;  
\item[$(c)$] If $x$ is an instance of $A$ and $R(x)$  is its corresponding instance of $B$, then
  there is some positive constant $\alpha$ such that 
  $\OPT_B(R(x)) \leq \alpha . \OPT_A(x)$~; 
\item[$(d)$] If $s$ is a solution of $R(x)$, then there is some positive constant $\beta$
  such that  \\ 
  $|\OPT_A(x) - c_A(S(s))| \leq \beta |\OPT_B(R(x)) - c_B(s)|$.  
\end{itemize} 

We prove Theorem \ref{theoreme1} by showing that the pair $(R,S)$
defined previously is an \emph{L-reduction} from \PB{Min-Vertex-Cover-3} to
\PB{EConsI} and from \PB{Min-Vertex-Cover-3} to \PB{EComI}. First note that
properties $(a)$ and $(b)$ are obviously satisfied by $R$ and $S$. 

Consider $\text{pb} \in \{\PB{EcomI}, \PB{EConsI}\}$. Let $G=(V,E)$ be a
cubic graph with $n$ vertices and $m$ edges. We now prove properties
$(c)$ and $(d)$. Consider the genomes $G_1$ and $G_2$ obtained by
$R(G)$. 
For sake of clarity, we abbreviate here and in the following
$\OPT_{\PB{Min-Vertex-Cover-3}}$ to $\OPT_{\PB{Min-VC}}$.
First, we need to prove that there exists $\alpha \geq 0$
such that $\OPT_{\text{pb}}(G_1,G_2) \leq \alpha . \OPT_{\PB{Min-Vertex-Cover-3}}(G)$.  

Since $G$ is cubic, we have the following properties:
\begin{align}
  & n \geq 4 \\
  & m = \frac{1}{2}\sum_{i=1}^n{degree(v_i)} = \frac{3 n}{2} \\
  & \OPT_{\PB{Min-VC}}(G) \geq \frac{m}{3} = \frac{n}{2}   
\end{align} 

To explain property (5), remark that, in a cubic graph $G$ with $n$
vertices and $m$ edges, each vertex covers three edges. Thus, a set of
$k$ vertices covers at most $3k$ edges. Hence, any vertex cover of $G$
must contain at least $\frac{m}{3}$ vertices.

By Lemma \ref{lemme3}, we know that sequences of the form $a_iC_if_i$,
$1 \leq i \leq n$, contain either zero or $W(\text{pb})$ non trivial
robust intervals. By Lemma \ref{lemme1}, there are no other non
trivial robust intervals. So, we have the following inequality:

\begin{center}
  $\OPT_{\text{pb}}(G_1,G_2) \leq \underbrace{T(\text{pb},G)}_{trivial ~robust
    ~intervals} + W(\text{pb})\cdot n$  
\end{center}

If $\text{pb} =$ \PB{EComI}, we have:
\begin{align}	
  \OPT_{\PB{EComI}}(G_1,G_2) \leq & ~7n + m + 3 + 4n \nonumber \\
  \OPT_{\PB{EComI}}(G_1,G_2) \leq & ~\frac{27 n}{2}  ~~by~(3) ~and~(4)   
\end{align}
And if $\text{pb} =$ \PB{EConsI}, we have :
\begin{align}
  \OPT_{\PB{EConsI}}(G_1,G_2) \leq & ~7n + m + 2 + n	\nonumber \\
  \OPT_{\PB{EConsI}}(G_1,G_2) \leq & ~\frac{21 n}{2}  ~~by~(3) ~and~(4) 
\end{align}

Altogether, by (5), (6) and (7), we prove property $(c)$ with $\alpha = 27$.

Now, let us prove property $(d)$. Let $VC = \{v_{i_1},v_{i_2} \ldots
v_{i_P}\}$ be a minimum vertex cover of $G$. Then 
$P = \OPT_{\PB{Min-VC}}(G)$. Let $G_1$ and $G_2$ be the genomes
obtained by $R(G)$. Let $(G_1,G_2^E)$ be an exemplarization of
$(G_1,G_2)$ and let $k'$ be the number of robust intervals of
$(G_1,G_2^E)$. Finally, let $VC'$ be the vertex cover of $G$ such that
$VC' = S(G_1,G_2^E)$.  We need to find a positive constant $\beta$
such that $|P - |VC'| | \leq \beta | \OPT_{\text{pb}}(G_1,G_2) - k'|$. 

For $\text{pb} \in \{\PB{EcomI}, \PB{EConsI}\}$, let $N_{\text{pb}}$ be the number
of robust intervals between the two genomes obtained by $F(VC)$.   
By the first property of Lemma \ref{lemme4}, we have
$$
\OPT_{\text{pb}}(G_1,G_2) \geq N_{\text{pb}} \geq  W(\text{pb})\cdot n + T(\text{pb},G)
-W(\text{pb})\cdot P
$$
So, it is sufficient to prove
that there exists some $\beta \geq 0$ such that $| P - |VC'| | \leq \beta |W(\text{pb})\cdot
n + T(\text{pb},G) -W(\text{pb})\cdot P - k'|$.
By the second property of Lemma \ref{lemme4}, we have $|VC'| = \frac{W(\text{pb})\cdot n + T(\text{pb},G) - k'}{W(\text{pb})}$. 
Since $P \leq |VC'|$, we have
$| P - |VC'| | = |VC'| - P = \frac{W(\text{pb})\cdot n + T(\text{pb},G) - k'}{W(\text{pb})}
- P = \frac{1}{W(\text{pb})}( W(\text{pb})\cdot n + T(\text{pb},G) -W(\text{pb})\cdot P-k')$. 

So $\beta = 1$ is sufficient in both cases, since $W(\PB{EComI})= 4$
and $W(\PB{EConsI}) = 1$, which implies $\frac{1}{W(\text{pb})} \leq 1$.  

Altogether, we then have 
$|\OPT_{\PB{Min-VC}}(G) - |VC'| | \leq 1 \cdot  
|\OPT_{\text{pb}}(G_1,G_2) - k'|$.

We proved that the reduction $(R,S)$ is an \emph{L-reduction}. This
implies that for two genomes $G_1$ and $G_2$, both problems
\PB{EConsI} and \PB{EComI} are \APXhard even if $\OCC(G_1) = 1$ and
$\OCC(G_2) = 2$. Theorem \ref{theoreme1} is proved.  \qed    

We extend in Corollary \ref{corollary_interval} our results for the
\emph{intermediate} and \emph{maximum matching} models. 

\begin{corollary}\label{corollary_interval}
  \PB{IComI}, \PB{MComI}, \PB{IConsI} and \PB{MConsI} are \APXhard even
  when genomes $G_1$ and $G_2$ are such that $\OCC(G_1) = 1$ and
  $\OCC(G_2) = 2$.  
\end{corollary}

\begin{proof}
  The \emph{intermediate} and  \emph{maximum} matching models are
  identical to the \emph{exemplar} model when one of the two genomes contains no
  duplicates. Hence, the \APXhardness result for \PB{EComI}
  (resp. \PB{EConsI}) also holds for \PB{IComI} and \PB{MComI}
  (resp. \PB{IConsI} and \PB{MConsI}). \qed    
\end{proof}


\section{\PB{EBD} is \APXhard}
\label{Hardness_breakpoints}

Consider two genomes $G_1$ and $G_2$ with duplicates, and let \PB{EBD}
(resp. \PB{IBD}, \PB{MBD}) be the problem which consists 
in finding  an exemplarization 
(resp. intermediate matching, maximum
matching) $(G'_1,G'_2,\mathcal{M})$ of $(G_1,G_2)$that minimizes the number of
breakpoints between $G'_1$ and 
$G'_2$. 

\PB{EBD} has been proved to be \NPC even if $\OCC(G_1) = 1$ and
$\OCC(G_2) = 2$~\cite{Complexite_exemplarisation}. Some inapproximability 
results also exist: in particular, it has been proved 
in~\cite{chen_inproceding_approximation_breakpoint} that, in the general case, 
\PB{EBD} cannot be approximated within a factor $c$ log $n$, where $c > 0$ is a 
constant, and cannot be approximated within a factor $1.36$ when 
$\OCC(G_1) = \OCC(G_2) = 2$. Moreover, for two balanced genomes $G_1$ and $G_2$ 
such that $k = \OCC(G_1) = \OCC(G_2)$, several approximation algorithms for \PB{MBD}
are given. These approximation algorithms admit respectively a ratio of 
$1.1037$ when $k=2$~\cite{breakpoint_approximation_kolmann}, $4$ when 
$k=3$~\cite{breakpoint_approximation_kolmann} and $4k$ in the general 
case~\cite{inproceeding_kolman_approximation_breakpoint}.
We can conclude from the above results that \PB{IBD} and \PB{MBD} problems
are also \NPC, since when one genome contains no duplicates,
\emph{exemplar}, \emph{intermediate} and \emph{maximum matching} models are equivalent.

In this section, we improve the above results by showing that the three
problems \PB{EBD}, \PB{IBD} and \PB{MBD} are \APXhard, even when
genomes $G_1$ and $G_2$ are such that $\OCC(G_1) = 1$ and $\OCC(G_2) =
2$. The main 
result is Theorem~\ref{theoremeEBD} below, which will be completed by Corollary~\ref{corollary_breakpoint}
at the end of the section.

\begin{theorem}\label{theoremeEBD}
  \PB{EBD} is \APXhard even when genomes $G_1$ and $G_2$ are such that $\OCC(G_1) = 1$ and $\OCC(G_2) = 2$. 
\end{theorem}

To prove Theorem \ref{theoremeEBD}, we use an \emph{L-Reduction} from
\PB{Min-Vertex-Cover-3} to \PB{EBD}. Let  $G = (V,E)$ be a
cubic graph with $V = \{v_1\ldots v_n\}$ and $E = \{e_1\ldots
e_m\}$. For each $i$, $1\leq i \leq n$, let $e_{f_i}$,
$e_{g_i}$ and $e_{h_i}$ be the three edges which are incident to $v_i$
in $G$ with $f_i < g_i < h_i$. Let $R'$ be the polynomial
transformation which associates to $G$  the following genomes $G_1$
and $G_2$, where each gene has a positive sign:  \\ 
$G_1 = a_0~a_1~b_1~a_2~b_2 \ldots a_n~b_n~c_1~d_1~c_2~d_2 \ldots
c_m~d_m~c_{m+1}$  \\ 
$G_2 = a_0~a_n~d_{f_n}~d_{g_n}~d_{h_n}~b_n\ldots
a_2~d_{f_2}~d_{g_2}~d_{h_2}~b_2
~a_1~d_{f_1}~d_{g_1}~d_{h_1}~b_1~c_1~c_2 \ldots c_m~c_{m+1}$ \\ 
with : 
\begin{itemize}
\item $a_0 = 0$, and for each $i$, $1 \leq i \leq n$, $a_i = i$ and $b_i = n+i$
\item $c_{m+1} = 2n + m + 1$, and for each $i$, $1 \leq i
  \leq m$, $c_i = 2n +i$ and $d_i = 2n + m + 1 + i$ 
\end{itemize}

We remark that there is no duplication in $G_1$, so $\OCC(G_1)=1$. In
$G_2$, only the genes $d_i$, $1 \leq i \leq m$, are
duplicated and occur twice. Thus  $\OCC(G_2) = 2$. 

Let $G$ be a cubic graph and $VC$ be a vertex cover of $G$. Let
$G_1$ and $G_2$ be the genomes obtained by $R'(G)$. We define $F'$ to
be the polynomial transformation which associates to $VC$, $G_1$
and $G_2$ the exemplarization $F'(VC)=(G_1,G_2^E)$ of $(G_1,G_2)$ as
follows. For each $i$ such that $v_i \notin VC$, we remove from $G_2$
the genes $d_{f_i}, d_{g_i}$ and $d_{h_i}$. Then, for each $j$, $1
\leq j \leq m$ such that $d_j$ still has two occurrences in
$G_2$, we arbitrarily remove one of these occurrences in order to
obtain the genome $G_2^E$. Hence, $(G_1,G_2^E)$ is an exemplarization
of $(G_1,G_2)$.  

Given a cubic graph $G$, we construct $G_1$ and $G_2$ by the
transformation $R'(G)$. Given an exemplarization $(G_1,G_2^E)$ of
$(G_1,G_2)$, let $S'$ be the polynomial transformation which
associates to $(G_1,G_2^E)$ the set $VC = \{ v_i | 1 \leq i
\leq n, a_i \text{~and~} b_i \text{~are not consecutive
  in~}G_2^E\}$. We claim that $VC$ is a vertex cover of
$G$. Indeed, let $e_p$, $1 \leq p \leq m$, be an edge of
$G$. Genome $G_2^E$ contains one occurrence of gene $d_p$ since
$G_2^E$ is an exemplarization of $G_2$. By construction, there exists
$i$, $1 \leq i \leq n$, such that $d_p$ is in
$G_2^E[a_i,b_i]$ and such that $e_p$ is incident to $v_i$. The
presence of $d_p$ in $G_2^E[a_i,b_i]$ implies that  vertex $v_i$
belongs to $VC$. We can conclude that each edge of $G$ is incident
to at least one vertex of $VC$.  

Lemmas \ref{lemmeEBD1} and \ref{lemmeEBD2} below are used to prove that $(R',S')$ is an \emph{L-Reduction} from the \PB{Min-Vertex-Cover-3} problem to the \PB{EBD} problem. Let $G=(V,E)$ be a cubic graph with $V=\{v_1,v_2 \ldots v_n\}$ and $E=\{e_1,e_2 \ldots e_m\}$ and let us construct $(G_1,G_2)$ by the transformation $R'(G)$.

\begin{lemma}  \label{lemmeEBD1}
  Let $VC$ be a vertex cover of $G$ and $(G_1,G_2^E)$ the
  exemplarization given by $F'(VC)$. Then $|VC| = k \Rightarrow
  B(G_1,G_2^E) \leq n + 2m + k + 1$, where $B(G_1,G_2^E)$ is the
  number of breakpoints between $G_1$ and $G_2^E$. 
\end{lemma}

\begin{proof}
  Suppose $|VC| = k$. Let us list the breakpoints between genomes $G_1$
  and $G_2^E$ obtained by $F'(VC)$. The pairs $(b_i,a_{i+1})$, $1
  \leq i \leq n-1$, and $(b_n,c_1)$ induce one breakpoint
  each. For all $i$, $1 \leq i \leq m$, each pair of the form
  $(c_i,d_i)$ (resp. $(d_i,c_{i+1})$) induces one breakpoint. For all $i$, $1
  \leq i \leq n$,  such that $v_i \in VC$, $(a_i,b_i)$ induces
  at most one breakpoint. Finally, the pair $(a_0,a_1)$ induces one
  breakpoint. Thus there are at most \textbf{$n + 2m + k + 1$}
  breakpoints of $(G_1,G_2^E)$.  \qed 
\end{proof}

\begin{lemma}  \label{lemmeEBD2}
  Let $(G_1,G_2^E)$ be an exemplarization of $(G_1,G_2)$ and $VC'$ be
  the vertex cover of $G$ obtained by $S'(G_1,G_2^E)$. We have
  $B(G_1,G_2^E) = k' \Rightarrow |VC'| = k'-n-2m-1$. 
\end{lemma}

\begin{proof}
  Let $(G_1,G_2^E)$ be an exemplarization of $(G_1,G_2)$ and $VC'$ be
  the vertex cover obtained by $S'(G_1,G_2^E)$. Suppose $B(G_1,G_2^E) =
  k'$. For any exemplarization $(G_1,G_2^E)$ of $(G_1,G_2)$, the
  following breakpoints always occur: the pair $(a_0,a_1)$~; for each
  $i$, $1 \leq i \leq m$, each pair $(c_i,d_i)$ and
  $(d_i,c_{i+1})$~; for each $i$, $1 \leq i \leq n-1$, the
  pair $(b_i,a_{i+1})$~; the pair $(b_n,c_1)$. Thus, we have at least $n
  + 2m + 1$ breakpoints. The other possible breakpoints are induced by
  pairs of the form of $(a_i,b_i)$. Since we have $B(G_1,G_2^E) = k'$,
  there are exactly $k' - n - 2m -1$ such breakpoints.  By construction
  of $VC'$, the cardinality of $VC'$ is equal to the number of
  breakpoints induced by pairs of the form $(a_i,b_i)$. So, we have:
  $|VC'| =  k'  - n - 2m -1$. \qed 
\end{proof}

To prove that $(R',S')$ is an {\it L-reduction}, we first notice that
properties {\it (a)} and {\it (b)} of an {\it L-reduction} are trivially verified.
The next lemma proves property {\it (c)}.

\begin{lemma}\label{lemmeEBD3}
  The inequality $\OPT_{\PB{EBD}}(G_1,G_2) \leq 12 \cdot \OPT_{\PB{Min-VC}}(G)$ holds. 
\end{lemma}

\begin{proof}
  For a cubic graph $G$ with $n$ vertices and $m$ edges, we have $2m =
  3n$ (see (4)) and  \linebreak $\OPT_{\PB{Min-VC}}(G)\geq \frac{n}{2}$(see
  (5)). By construction of the genomes $G_1$ and $G_2$, any
  exemplarization of $(G_1,G_2)$ contains $2n+2m+2$ genes in each
  genome. Thus, we have $\OPT_{\PB{EBD}}(G_1,G_2) \leq 2n + 2m
  +2\leq 6n$ ($n\geq 4$ in a cubic graph). Hence,  we conclude that $\OPT_{\PB{EBD}}(G_1,G_2)
  \leq 12 \cdot  \OPT_{\PB{Min-VC}}(G)$. \qed 
\end{proof}

Now, we prove property ${\it (d)}$ of our {\it L-reduction}.

\begin{lemma}\label{lemmeEBD4}
  Let $(G_1,G_2^E)$ be an exemplarization of $(G_1,G_2)$ and let $VC'$
  be the vertex cover of $G$ obtained by $S'(G_1,G_2^E)$. Then, we have
  $|\OPT_{\PB{Min-VC}}(G)  - |VC'| | \leq | \OPT_{\PB{EBD}}(G_1,G_2) -
  B(G_1,G_2^E)|$  
\end{lemma}

\begin{proof}
  Let $(G_1,G_2^E)$ be  an exemplarization of $(G_1,G_2)$ and $VC'$ be
  the vertex cover of $G$ obtained by $S'(G_1,G_2^E)$. Let $VC$ be a
  vertex cover of $G$ such that $|VC| = \OPT_{\PB{Min-VC}}(G)$.  

  We know that $\OPT_{\PB{Min-VC}}(G) \leq |VC'|$ and
  $\OPT_{\PB{EBD}}(G_1,G_2) \leq B(G_1,G_2^E)$. So, it is sufficient
  to prove $|VC'| - \OPT_{\PB{Min-VC}}(G) \leq B(G_1,G_2^E) -
  \OPT_{\PB{EBD}}(G_1,G_2)$.  

  By Lemma \ref{lemmeEBD1}, we have $B(F'(VC)) \leq n + 2m + 1 +
  \OPT_{\PB{Min-VC}}$,  which implies $\OPT_{\PB{EBD}}(G_1,G_2) \leq
  B(F'(VC)) \leq n + 2m + 1 + \OPT_{\PB{Min-VC}}$. Then   
  \begin{align}
    B(G_1,G^E_2) - \OPT_{EBD}(G_1,G_2) \geq B(G_1,G_2^E) - n -2m -1 -\OPT_{\PB{Min-VC}}(G) 
  \end{align}
  By Lemma \ref{lemmeEBD2}, we have: $|VC'|=B(G_1,G_2^E)-n-2m-1$ which implies 
  \begin{align}
    |VC'| - \OPT_{\PB{Min-VC}}(G) = B(G_1,G_2^E)-n-2m -1 - \OPT_{\PB{Min-VC}}(G)
  \end{align}
  Finally, by (8) and (9), we get $|VC'| - \OPT_{\PB{Min-VC}} \leq
  B(G_1,G_2^E) - \OPT_{\PB{EBD}}(G_1,G_2)$. \qed 
\end{proof}

Lemmas \ref{lemmeEBD3} and \ref{lemmeEBD4} prove that the pair
$(R',S'\begin{scriptsize}\end{scriptsize})$ is an \emph{L-reduction}
from \PB{Min-Vertex-Cover-3} to \PB{EBD}. Hence, \PB{EBD} is \APXhard even if
$\OCC(G_1) = 1$ and $\OCC(G_2) = 2$, and Theorem \ref{theoremeEBD} is
proved. We extend in Corollary \ref{corollary_breakpoint} our results
for the \emph{intermediate} and \emph{maximum matching} models.  

\begin{corollary}\label{corollary_breakpoint}
  The \PB{IBD} and \PB{MBD} problems are \APXhard even when 
  genomes $G_1$ and $G_2$ are such that $\OCC(G_1) = 1$ and $\OCC(G_2) = 2$.  
\end{corollary} 

\begin{proof}
  The intermediate and maximum matching models are identical to
  the exemplar model when one of the two genomes contains no duplicates. Hence, the
  \APXhardness result for \PB{EBD} also holds for \PB{IBD} and
  \PB{MBD}. \qed
\end{proof}

\section{Zero breakpoint distance}
\label{Complexity_ZXBD}

This section is devoted to zero breakpoint distance recognition
issues.
Indeed, in~\cite{chen_inproceding_approximation_breakpoint}, the authors showed that deciding whether the
exemplar breakpoint distance between any two genomes is zero or not is
\NPC even when no gene occurs more than three times in both
genomes, \emph{i.e.}, instances of type $(3,3)$. 
This important result implies that the exemplar breakpoint distance
problem does not admit any approximation in polynomial-time, unless
$\Pclass = \NPclass$.
Following this line of research, we first complement the result of
\cite{chen_inproceding_approximation_breakpoint} by proving that deciding whether the exemplar
breakpoint distance between any two genomes is zero or not is
\NPC, even when no gene is duplicated more than twice in one of the genomes 
(the maximum number of duplications is however unbounded in the other
genome). 
This result is next extended to the intermediate matching model and we
give a practical -~but exponential~- algorithm for deciding whether
the exemplar breakpoint distance between any two genomes is zero or
not in case no gene occurs more than twice in both genomes 
(a problem whose complexity, \Pclass \emph{versus} \NPC, remains open).
Finally, we show that deciding whether the maximum matching 
breakpoint distance between any two genomes is zero or not is
polynomial-time solvable and hence that such negative approximation
results (the ones we obtained for the exemplar and intermediate
models) do no propagate to the maximum matching model. 

The following easy observation will prove extremely useful in
the sequel of the present section. 

\begin{observation}
  \label{obs:zero}
  Let $G_1$ and $G_2$ be two genomes.
  If the exemplar breakpoint distance between $G_1$ and $G_2$ is zero,
  then there exists an exemplarization $(G^E_1, G^E_2)$ of
  $(G_1, G_2)$ such that
  (1)
  $G^E_1 = G^E_2$, or 
  (2)
  $-(G^E_1)^r = G^E_2$, where $-(G^E_1)^r$ is the signed reversal of
  genome $G_1$.
  The same observation can be made for the intermediate and maximum
  matching models. 
\end{observation}

\subsection{Zero exemplar breakpoint distance}
\label{section:ZEBD}


The zero exemplar breakpoint distance (\PB{ZEBD}) problem is formally
defined as follows. 

\medskip
\mabox{
  \textbf{Problem:} \PB{ZEBD}\\
  \textbf{Input:} Two genomes $G_1$ and $G_2$. \\
  \textbf{Question:} Is the exemplar breakpoint distance between $G_1$
  and $G_2$ equal to zero?
}
\medskip

Aiming at precisely defining the inapproximability landscape of
computing the exemplar breakpoint distance between two genomes,
we complement the result of
\cite{chen_inproceding_approximation_breakpoint}, who showed \PB{ZEBD}
to be \NPC even for instances of type $(3,3)$, by the following
theorem.  

\begin{theorem}
  \label{theorem:ZEBD_2k}
  \PB{ZEBD} is \NPC even if no gene
  occurs more than twice in $G_1$. 
\end{theorem}

\begin{proof}
  Membership of \PB{ZEBD} to \NPclass is immediate.
  The reduction we use to prove hardness is from \PB{Min-Vertex-Cover}~\cite{Garey:Johnson:1979}. 
  Let an arbitrary instance of \PB{Min-Vertex-Cover} be given by a graph
  $G = (V,E)$ and a positive integer $k$.
  Write $V = \{v_1, v_2 \ldots v_n\}$ and
  $E = \{e_1, e_2 \ldots e_m\}$.
  In the rest of the proof, elements of $V$ (resp. $E$) will be seen 
  either as vertices (resp. edges) or genes, depending on the context.
  The corresponding instance $(G_1, G_2)$  of \PB{ZEBD} is defined as
  follows: 
  \begin{align*}
    G_1 &= v_1 ~ X_1 ~ v_2 ~ X_2 \ldots v_n ~ X_n \\
    G_2 &= Y[1] ~ Y[2] ~ \ldots ~ Y[k] ~ Y_V\text{.}
  \end{align*} 
  For each $i = 1, 2, \ldots, n$, $X_i$ is defined to be
  $X_i = e_{i_1} ~ e_{i_2} ~ \ldots ~ e_{i_j}$,
  where $e_{i_1}, e_{i_2}, \ldots, e_{i_j}$,
  $i_1 < i_2 < \ldots < i_j$, are the edges incident to
  vertex $v_i$. The strings $Y[i]$, $1 \leq i \leq k$, are all equal and 
  are defined by $Y[i] = Y_V ~ Y_E$ where
  $Y_V = v_1 ~ v_2 ~ \ldots ~ v_n$ and
  $Y_E = e_1 ~ e_2 ~ \ldots ~ e_m$.

  Notice that no gene occurs more than twice in $G_1$ (actually
  genes $v_i$ occur once and genes $e_i$ occur twice). However, the
  number of occurrences of each gene in $G_2$ is upper bounded by
  $k+1$. 
  Furthermore, all genes have positive sign, and hence according to
  Observation~\ref{obs:zero} we only need to consider exemplarizations
  $(G^E_1, G^E_2)$ of $(G_1, G_2)$ such that $G^E_1 = G^E_2$.

  It is immediate to check that our construction can be carried out in
  polynomial-time.
  We now claim that there exists a vertex cover of size $k$ in $G$ iff
  the exemplar breakpoint distance between $G_1$ and $G_2$ 
  is zero. 

  Suppose first that there exists a vertex cover $V' \subseteq V$
  of size $k$ in $G$. 
  Write $V' = \{v_{i_1}, v_{i_2}, \ldots, v_{i_k}\}$,
  $i_1 < i_2 < \ldots < i_k$.
  For convenience, we also define $i_0$ to be $0$.
  From $V'$ we construct an exemplarization $(G^E_1,G^E_2)$ as follows. 
  We obtain $G^E_1$ from $G_1$ by a two step procedure.
  First we delete in $G_1$ all strings $X_i$ such that $v_i \notin V'$. 
  Second, for each $1 \leq j \leq m$, if gene $e_j$ still occurs
  twice, we delete its second occurrence (this second step is concerned with
  edges connecting two vertices in $V'$). 
  We now turn to $G^E_2$.
  For $1 \leq j \leq k$, we consider the string
  $Y[j] = Y_V ~ Y_E$ that we process as follows:
  (1) we delete in $Y_V$ all genes but $v_{i_j}$ and
  those genes $v_\ell \notin V'$ such that $i_{j-1} < \ell < i_j$, and 
  (2) we delete in $Y_E$ all genes but those $e_\ell$ that are not
  incident to $v_{i_j}$ or incident to $v_{i_j}$ and some smaller
  vertex in $V'$ (\emph{i.e.}, $e_\ell = \{v_{i_{j'}}, v_{i_j}\}$ for
  some $j' < j$). 
  Finally, we delete in the trailing string 
  $Y_V = v_1 ~ v_2 ~ \ldots ~ v_n$
  all genes but those $v_\ell$ ($\notin V'$) such that $i_k < \ell$.
  Since $V'$ is a vertex cover in $G$, then it follows that each gene
  occurs once in the obtained genomes,
  \emph{i.e.}, $(G^E_1, G^E_2)$ is indeed an exemplarization of 
  $(G_1, G_2)$.
  It is now easily seen that $G^E_1 = G^E_2$, and hence that the
  exemplar breakpoint distance between $G_1$ and $G_2$ is zero.

  Conversely, suppose that the exemplar breakpoint distance between
  $G_1$ and $G_2$ is zero. Since all genes have a positive sign, then it
  follows that there exists an exemplarization $(G^E_1, G^E_2)$ of 
  $(G_1, G_2)$ such that $G^E_1 = G^E_2$.
  Exemplarization $G^E_2$ can be written as
  $$
  G^E_2 = 
  Y_V[1] ~ Y_E[1] ~ Y_V[2] ~ Y_E[2] ~ \ldots Y_V[k] ~ Y_E[k] ~ Y_V[k+1]
  $$
  where, $Y_V[i]$, $1 \leq i \leq k+1$, is a string on $V$ 
  and $Y_E[i]$, $1 \leq i \leq k$, is a string on $E$,  
  $V$ and $E$ being viewed as alphabets. 
  Now, define $V' \subseteq V$ as follows: $v_i \in V'$ iff
  gene $v_i$ occurs in some $Y_V[j]$, $1 \leq j \leq k$, as the last gene.
  By construction, $|V'| \leq k$ (we may indeed have $|V'| < k$ if some 
  $Y_V[j]$, $1 \leq j \leq k$, denotes the empty string).
  We now observe that, since no gene $v_i$ is duplicated in $G_1$, all
  genes $e_\ell$ that occur between some gene $v_i \in V'$ and some
  gene $v_j \in V$ in $G^E_2$ should match genes in string $X_i$ in
  $G_1$. Then it follows that $V'$ is a vertex cover of size at
  most $k$ in $G$. 
  \qed
\end{proof}

The complexity of \PB{ZEBD} remains open in case no gene occurs
more than twice in $G_1$ and more than a constant times in $G_2$,
\emph{i.e.}, instances of type $(2, c)$ for some $c = O(1)$~;
recall here that \PB{ZEBD} is \NPC if no gene occurs more than three
times in $G_1$ or in $G_2$ (instances of type $(3,3)$, \cite{chen_inproceding_approximation_breakpoint}).
In particular, the complexity of \PB{ZEBD} for instances of type
$(2,2)$ is open.
However, we propose here a practical -~but exponential~- algorithm for
\PB{ZEBD} for instances of type $(2,2)$, which is well-suited in case the
number of genes that occur twice both in $G_1$ and in $G_2$ is relatively
small. 

\begin{proposition}\label{prop:ZEBD_22}
  \PB{ZEBD} for instances of type $(2,2)$ (no gene occurs more than
  twice in $G_1$ and in $G_2$) is solvable in $O^*(1.6182^{2k})$
  time, where $k$ is upper-bounded by the number of genes that occur
  exactly twice in $G_1$ and in $G_2$.
\end{proposition}

\begin{proof}
  According to Observation~\ref{obs:zero}, for any instance $(G_1,
  G_2)$, we only need to focus on exemplarizations $(G^E_1, G^E_2)$
  such that $G^E_1 = G^E_2$ or $-(G^E_1)^r = G^E_2$, where $-(G^E_1)^r$
  is the signed reversal of $G^E_1$.
  Let us first consider the case $G^E_1 = G^E_2$ (the case 
  $-(G^E_1)^r = G^E_2$ is identical up to a signed reversal and will
  thereby be briefly discussed at the end of the proof).

  Let $(G_1, G_2)$ be an instance of type $(2, 2)$ of
  \PB{ZEBD}.
  Our algorithm is by transforming instance $(G_1, G_2)$ into 
  a CNF boolean formula $\phi$ with only few large clauses such that
  $\phi$ is satisfiable iff the exemplar breakpoint
  distance between $G_1$ and $G_2$ is zero.
  By hypothesis, each signed gene occurs at most twice in $G_1$
  and in $G_2$.
  Therefore, for any signed gene $g$, we have one out of four
  possible distinct configurations depicted in
  Figure~\ref{fig:ZEBD_22},  
  where $p_1$, $p_2$, $q_1$ and $q_2$ are positions of occurrence of
  $g$ in $G_1$ and $G_2$.
  Furthermore, since we are looking for an exemplarization 
  $(G^E_2, G^E_2)$ of $(G_1, G_2)$ such that $G^E_1 = G^E_2$, we may
  assume, in case $g$ occurs only once in $G_1$ or in $G_2$, that all
  occurrences of $G$ have the same sign (otherwise a trivial
  self-reduction would indeed apply). In other words, referring 
  at Figure~\ref{fig:ZEBD_22}, we assume
  $G_1[p_1] = G_2[q_1] = G_2[q_2]$ in case (2),
  $G_1[p_1] = G_1[p_2] = G_2[q_1]$ in case (3), and
  $G_1[p_1] = G_2[q_1]$ in case (4).
  Finally, as for case (1), we may assume that either all occurrences
  have the same sign,
  or $G_1[p_1] = -G_1[p_2]$ and $G_2[q_1] = -G_2[q_2]$
  (otherwise a trivial self-reduction would again apply).

  \begin{figure}
    \begin{center}
      \input{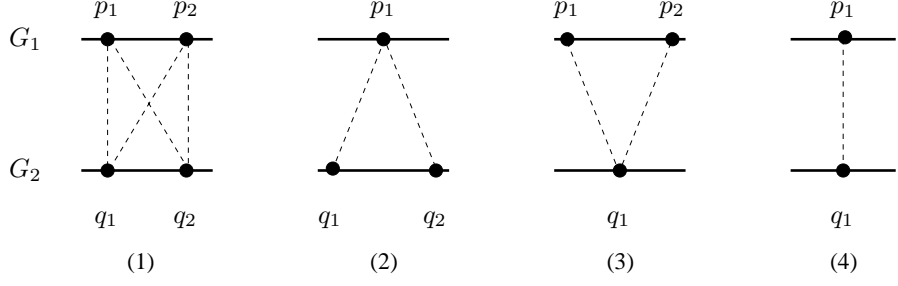}
    \end{center}
    \caption{\label{fig:ZEBD_22}
      The $4$ gene-configurations for instances of type $(2,2)$:
      $p_1$ and $p_2$ are the occurrence positions of gene $g$ in $G_1$,
      and $q_1$ and $q_2$ are the occurrence positions of gene $g$ in
      $G_2$.} 
  \end{figure} 

  We now describe the construction of the CNF boolean
  formula $\phi$.
  First, the set of boolean variables $X$ is defined as follows:
  for each gene $g$ occurring at position $p$ in $G_1$ and at position
  $q$ in $G_2$ (\emph{i.e.}, $|G_1[p]| = |G_2[q])|$) we add to $X$ the
  boolean variable $x^p_q$. 
  We now turn to defining the clauses of $\phi$.
  Let $g$ be any gene, and let the occurrence positions of $g$ in $G_1$
  and in $G_2$ be noted as in Figure~\ref{fig:ZEBD_22}.

  \begin{itemize}

  \item 
    if $\OCC(g, G_1) = \OCC(g, G_2) = 2$ (case(1)),
    \begin{itemize}
    \item[--]
      if $G_1[p_1] = G_1[p_2] = G_2[q_1] = G_2[q_2]$,
      we add to $\phi$ the clauses
      $(x^{p_1}_{q_1} \vee x^{p_1}_{q_2} \vee x^{p_2}_{q_1}\vee x^{p_2}_{q_2})$,
      $(\overline{x^{p_1}_{q_1}} \vee \overline{x^{p_1}_{q_2}})$,
      $(\overline{x^{p_1}_{q_1}} \vee \overline{x^{p_2}_{q_1}})$,
      $(\overline{x^{p_1}_{q_1}} \vee \overline{x^{p_2}_{q_2}})$,
      $(\overline{x^{p_1}_{q_2}} \vee \overline{x^{p_2}_{q_1}})$,
      $(\overline{x^{p_1}_{q_2}} \vee \overline{x^{p_2}_{q_2}})$ and
      $(\overline{x^{p_2}_{q_1}} \vee \overline{x^{p_2}_{q_2}})$,

    \item[--]
      otherwise, we have $G_1[p_1] = -G_1[p_2]$ and 
      $G_2[q_1] = -G_2[q_2]$ (see above discussion),
      \begin{itemize}
      \item[--]
        if $G_1[p_1] = G_2[q_1]$ and $G_1[p_2] = G_2[q_2])$), 
        we add to $\phi$ the clauses
        $(x^{p_1}_{q_1} \vee x^{p_2}_{q_2})$ and
        $(\overline{x^{p_1}_{q_1}} \vee \overline{x^{p_2}_{q_2}})$,
      \item[--]
        if $G_1[p_1] = G_2[q_2]$ and $G_1[p_2] = G_2[q_1])$), 
        we add to $\phi$ the clauses
        $(x^{p_1}_{q_2} \vee x^{p_2}_{q_1})$ and
        $(\overline{x^{p_1}_{q_2}} \vee \overline{x^{p_2}_{q_1}})$,
      \end{itemize}
    \end{itemize}

  \item 
    if $\OCC(g, G_1) = 1$ and  $\OCC(g, G_2) = 2$ (case (2)),
    we add to $\phi$
    the clauses $(x^{p_1}_{q_1} \vee x^{p_1}_{q_2})$ and 
    $(\overline{x^{p_1}_{q_1}} \vee \overline{x^{p_1}_{q_2}})$,

  \item
    if $\OCC(g, G_1) = 2$ and  $\OCC(g, G_2) = 1$ (case (3)), 
    we add to $\phi$
    the clauses $(x^{p_1}_{q_1} \vee x^{p_2}_{q_1})$ and 
    $(\overline{x^{p_1}_{q_1}} \vee \overline{x^{p_2}_{q_1}})$, and

  \item 
    if $\OCC(g,G_1) = \OCC(g,G_2) = 1$ (case (4)), we add to
    $\phi$ the clause $(x^{p_1}_{q_1})$.

  \end{itemize}
  
  The rationale of this construction is that if formula $\phi$
  evaluates to true for some assignment $f$ and $f(x^p_q)$ is true for
  some gene $g$ occurring at position $p$ in $G_1$ and $q$ in $G_2$,
  then all occurrences of $g$ but the one at position $p$ should be
  deleted in $G_1$ 
  and
  all occurrences of $g$ but the one at position $q$ should be deleted
  in $G_2$, in order to obtain the exemplar solution.
  What is left is to enforce that $\phi$ evaluates to true iff
  the exemplar breakpoint distance between $G_1$ and $G_2$ is zero.
  To this aim, we add to $\phi$ the following clauses.
  For each pair of variables $(x^{i_1}_{j_1}, x^{i_2}_{j_2})$ such
  that $|G_1[i_1]| \neq |G_1[i_2]|$,
  $i_1 < i_2$ and $j_1 > j_2$,
  we add to $\phi$ the clause 
  $(\overline{x^{i_1}_{j_1}} \vee \overline{x^{i_2}_{j_2}})$.
  The construction of $\phi$ is now complete.
  
  Clearly, $\phi$ evaluates to true iff 
  the exemplar breakpoint distance between $G_1$ and $G_2$ is zero.
  Let $k$ be the number of genes $g$ that occur twice in $G_1$ and in
  $G_2$ with the same sign, 
  \emph{i.e.}, $G_1[p_1] = G_1[p_2] = G_2[q_1] = G_2[q_2]$.
  We now make the important observation that all clauses in $\phi$
  have size less than or equal to $2$ except those $k$ clauses of size $4$ introduced in
  case gene $g$ occurs twice in $G_1$ and in $G_2$ with the same sign.
  By introducing a new boolean variable, we can easily replace in
  $\phi$ each clause of size $4$ by two clauses of size $3$,
  and hence we may now assume that $\phi$ is a $3$-CNF formula 
  (\emph{i.e.}, each clause has size at most $3$) with exactly $2k$ clauses of
  size $3$.   

  As for the case $-(G^E_1)^r = G^E_2$, we replace $G_1$ by $-(G_1)^r$
  and construct another $3$-CNF formula $\phi '$ as described above.
  The two $3$-CNF formulas need, however, to be examined separately.

  Fernau proposed in~\cite{Fernau} an algorithm for solving $3$-CNF
  boolean formulas that runs in $O^*(1.6182^{\ell})$ time, where $\ell$
  is the number of clauses of size $3$.
  Therefore, \PB{ZEBD} for instances of type $(2, 2)$ is solvable in
  $O^*(1.6182^{2k})$ time, where $k$ is the number of genes $g$ that
  occur twice in $G_1$ and in $G_2$. 
  \qed
\end{proof}


\subsection{Zero intermediate matching breakpoint distance}

We now turn to the zero intermediate breakpoint distance (\PB{ZIBD})
problem.  It is defined as follows.

\medskip
\mabox{
  \textbf{Problem:} \PB{ZIBD}\\
  \textbf{Input:} Two genomes $G_1$ and $G_2$. \\
  \textbf{Question:} Is the intermediate breakpoint distance between
  $G_1$ and $G_2$ equal to zero ?
}
\medskip

We show here that \PB{ZEBD} and \PB{ZIBD} are equivalent problems.
We need the following lemma.

\begin{lemma}[\cite{Angibaud:Fertin:Rusu:Thevenin:Vialette:RCG07Journal}]
  \label{lemma:eplucher}
  Let $G_1$ and $G_2$ be two genomes without duplicates and with the
  same gene content, and $G'_1$ and $G'_2$ be the two genomes obtained
  from $G_1$ and $G_2$ by deleting any gene $g$. 
  Then $B(G'_1, G'_2) \leq B(G_1, G_2)$.
\end{lemma}

\begin{theorem} 
  \label{theorem:ZIBD}
  \PB{ZEBD} and \PB{ZIBD} are equivalent problems. 
\end{theorem}
\begin{proof}
  One direction is trivial (any exemplarization is indeed an intermediate
  matching).
  The other direction follows from Lemma~\ref{lemma:eplucher}.
  \qed
\end{proof}

It follows from Theorem~\ref{theorem:ZIBD} that the problem \PB{IBD} is not
approximable even for instances of type $(3,3)$ (see~\cite{chen_inproceding_approximation_breakpoint}) and
if no gene occurs more than twice in $G_1$
(see Theorem~\ref{theorem:ZEBD_2k}). 

\subsection{Zero maximum matching breakpoint distance}

We show here that, oppositely to the exemplar and the intermediate
matching models,
deciding whether the maximum matching breakpoint distance between two
genomes is equal to zero is polynomial-time solvable, and hence we cannot rule
out the existence of accurate approximation algorithms for the maximum
matching model. We refer to this problem as \PB{ZMBD}.

\medskip
\mabox{
  \textbf{Problem:} \PB{ZMBD}\\
  \textbf{Input:} Two genomes $G_1$ and $G_2$. \\
  \textbf{Question:} Is the maximum matching breakpoint distance
  between $G_1$ and $G_2$ equal to zero ?
}
\medskip

The main idea of our approach is to transform any instance of
\PB{ZMBD} into a \emph{matching diagram} and next use an efficient
algorithm for finding a large set of non-intersecting line segments.
Note that this latter problem is equivalent to finding
a large increasing subsequence in permutations.

A matching diagram \cite{Golumbic:1980} consists of, say $n$, points
on each of two parallel lines, and $n$ straight line segments matching
distinct pairs of points. 
The intersection graph of the line segments is called a 
\emph{permutation graph}
(the reason for the name is that if the points on the top line are
numbered $1, 2, \ldots, n$, then the points on the other line are
numbered by a permutation on $1, 2, \ldots, n$).

We describe how to turn the pair of genomes $(G_1,G_2)$ into a
matching diagram $D(G_1, G_2)$.
For sake of presentation we introduce the following notations.
For each gene family $g$, we write $\OCC_{\text{pos}}(G, g)$
(resp. $\OCC_{\text{neg}}(G, g)$) for the number of positive (resp. negative) 
occurrences of gene $g$ in genome $G$.
According to Observation~\ref{obs:zero}, it is enough to consider two
cases: $G^M_1 = G^M_2$ or $-(G^M_1)^r = G^M_2$, where 
($G^M_1, G^M_2,\mathcal{M})$ is a maximum matching of $(G_1, G_2)$.

Let us first focus on testing $G^M_1 = G^M_2$
(the case $-(G^M_1)^r = G^M_2$ is identical up to a signed reversal).
We describe the construction of the top labeled points.
Reading genome $G_1$ from left to right, we replace gene $g$ 
by the sequence of labeled points 
$$
+\GP_1(i, \OCC_{\text{pos}}(G_2, g)) \;\; 
+\GP_1(i, \OCC_{\text{pos}}(G_2, g)-1) \;\;
\ldots \;\;
+\GP_1(i, 1)
$$
if $g$ is the $i$-th positive occurrence of gene $g$ in genome $G_1$
or  by the sequence of labeled points 
$$
-\GP_1(i, \OCC_{\text{neg}}(G_2, g)) \;\; 
-\GP_1(i, \OCC_{\text{neg}}(G_2, g)-1) \;\;
\ldots \;\;
-\GP_1(i, 1)
$$
if $g$ is the $i$-th negative occurrence of gene $g$ in genome $G_1$. 
A symmetric construction is performed for the labeled points of the
bottom line, \emph{i.e.},
reading genome $G_2$ from left to right, we replace gene $g$ by 
the sequence of labeled points 
$$
+\GP_2(i, \OCC_{\text{pos}}(G_1, g)) \;\; 
+\GP_2(i, \OCC_{\text{pos}}(G_1, g)-1) \;\;
\ldots \;\;
+\GP_2(i, 1)
$$
if $g$ is the $i$-th positive occurrence of gene $g$ in genome $G_2$
or  by the sequence of labeled points 
$$
-\GP_2(i, \OCC_{\text{neg}}(G_1, g)) \;\; 
-\GP_2(i, \OCC_{\text{neg}}(G_1, g)-1) \;\;
\ldots \;\;
-\GP_2(i, 1)
$$
if $g$ is the $i$-th negative occurrence of gene $g$ in genome $G_2$. 
We now obtain the matching diagram $D(G_1, G_2)$ as follows:
each labeled point $+\GP_1(i, j)$ 
(resp. $-\GP_1(i, j)$) of the top line
is connected to the labeled point $+\GP_2(j, i)$
(resp. $-\GP_2(j, i)$) of the bottom line by a line segment.
Clearly, each labeled point is incident to exactly one line segment,
and hence $D(G_1, G_2)$ is indeed a matching diagram.

Of particular importance, observe that by construction, for any 
$x \in \{1, 2\}$ and any two labeled points 
$+\GP_x(i,j)$ and $+\GP_x(i, k)$, $j \neq k$, the two line segments
incident to these two points are intersecting~; 
the same conclusion can be drawn for any two labeled points
$-\GP_x(i,j)$ and $-\GP_x(i, k)$, $j \neq k$.
The following lemma states this property in a suitable way.

\begin{lemma}
  \label{lemma:matching diagram}
  If
  $[+\GP_1(i, j), +\GP_2(j, i)]$ 
  and
  $[+\GP_1(k, \ell), +\GP_2(\ell, k)]$
  (resp. 
  $[-\GP_1(i, j), -\GP_2(j, i)]$ 
  and \linebreak
  $[-\GP_1(k, \ell), -\GP_2(\ell, k)]$)
  are two non-intersecting line segments in the matching diagram 
  $D(G_1, G_2)$, then 
  $i \neq k$ and $j \neq \ell$.
\end{lemma}

\begin{theorem} \label{theorem:ZMBD}
  \PB{ZMBD} is polynomial-time solvable.
\end{theorem}

\begin{proof}
  Let $G_1$ and $G_2$ be two genomes, and $m$ the size of a maximum
  matching between $G_1$ and $G_2$.
  According to Lemma~\ref{lemma:matching diagram}, there exists a
  maximum matching $(G^M_1, G^M_2,\mathcal{M})$ of $(G_1, G_2)$ such that 
  $G^M_1 = G^M_2$ if there exists $m$ non-intersecting line segments
  in $D(G_1, G_2)$.
  The maximum number of non-intersecting line segments in a matching
  diagram with $n$ points on each line can be found in
  $O(n \log \log n)$ time \cite{Chang:Wang:IPL:92}.

  As for the case $-(G^M_1)^r = G^M_2$, we replace $G_1$ by $-(G_1)^r$
  and run the same algorithm on the obtained matching diagram. 
  \qed
\end{proof}

\section{Approximating the number of adjacencies in the maximum matching model}
\label{Approximation_adjacences}

For two balanced genomes $G_1$ and $G_2$, several approximation
algorithms for computing the number of \emph{breakpoints} between
$G_1$ and $G_2$ are given for the maximum matching
model~\cite{breakpoint_approximation_kolmann,inproceeding_kolman_approximation_breakpoint}.
We propose in this section three approximation algorithms to maximize
the number of \emph{adjacencies} (as opposed to minimizing the number of
breakpoints). 
The approximation ratios we obtain are 
$1.1442$ when $\OCC(G_1) = 2$, 
$3$ when $\OCC(G_1) = 3$ and 
$4$ in the general case.
Observe that in the latter case, oppositely
to~\cite{breakpoint_approximation_kolmann,inproceeding_kolman_approximation_breakpoint},  
our approximation ratio is independent of the maximum number of
duplicates. 
Note also that in \cite{adjacence_chen}, inapproximation
results are given for two {\em unbalanced} genomes $G_1$ and $G_2$ even when
$\OCC(G_1) = 1$ and $\OCC(G_1) = 2$.

We first define the problem \PB{Max-$k$-Adj} we are interested in
($k \geq 1$ is a fixed integer).

\medskip
\mabox{\textbf{Problem:} \PB{Max-$k$-Adj} \\
  \textbf{Input:} Two balanced genomes $G_1$ and $G_2$ with $\OCC(G_1) = k$ (and consequently $\OCC(G_2) = k$). \\
  \textbf{Solution:} A maximum matching $(G_1^M,G_2^M,\mathcal{M})$ of $(G_1,G_2)$.\\
  \textbf{Measure:} The number of adjacencies between $G_1^M$ and $G_2^M$. 
}
\medskip

We define \PB{Max-Adj} to be the problem \PB{MAX~$k$-Adj}, in which
$k$ is unbounded. 

\subsection{A $1.1442$-approximation for \PB{Max-2-Adj}}

We focus here on balanced genomes $G_1$ and $G_2$ such that 
$\OCC(G_1) = 2$, and we give an approximation algorithm for
\PB{Max-2-Adj} based on the \PB{Max-2-CSP} problem
(defined below), for which a $1.1442$-approximation 
algorithm is given in \cite{approx_CSP}. 
The main idea is to construct a boolean formula $\varphi$ for each
possible adjacency, and next to maximize the number of boolean
formulas $\phi$ that can be simultaneously satisfied in a truth
assignment~;
the number of simultaneously satisfied formulas 
will be exactly the number of adjacencies, and hence
any approximation ratio for \PB{Max-2-CSP} is an approximation ratio
for \PB{Max-2-Adj}.

\medskip
\mabox{\textbf{Problem:} \PB{Max-$k$-CSP}  \\
  \textbf{Input:} A pair $(\chi,\Phi)$, where $\chi$ is a set of
  boolean variables and $\Phi$ is a set of boolean formulas such
  that each formula contains at most $k$ literals of $\chi$. \\
  \textbf{Solution:} An assignment of $\chi$. \\
  \textbf{Measure:} The number of formulas that are satisfied by the
  assignment.  
}
\medskip

We define the following transformation \ALGO{MakeCSP} that associates to
any instance of \PB{Max-2-Adj} an instance of \PB{Max-2-CSP}. 
Given
an instance $(G_1,G_2)$ of \PB{Max-2-Adj}, we create a variable $X_g$
for each gene $g$ and define $\chi$ as the set of variables
$X_g$. Then, we construct the set $\Phi$ of formulas. For each duo $d_i
= (G_1[i],G_1[i+1])$, $1 \leq i \leq n_{G_1} - 1$, such that
$d_i$ or $-d_i$ appears in $G_2$, we distinguish three cases
in order to create a formula $\varphi_i$ of $\Phi$:   

\begin{enumerate}
\item There exists a unique duo $d_j = (G_2[j],G_2[j+1])$ in $G_2$
  such that $d_j = d_i$ or $d_j = -d_i$. For sake of readability, we define the
  literal $Y_p^q$, $1\leq p\leq n_{G_1}$, $1\leq
  q\leq n_{G_2}$, where $|G_1[p]| = |G_2[q]|$, as follows: $Y_p^q =
  X_{|G_1[p]|}$ if $N_{G_1}[p] = N_{G_2}[q]$ and $Y_p^q =
  \overline{X_{|G_1[p]|}}$ otherwise. We now consider two cases: 

  \begin{itemize}
  \item(a) $d_i = d_j$: in that case, $\varphi_i = (Y_i^j \wedge Y_{i+1}^{j+1})$.
  \item(b) $d_i = -d_j$: in that case, $\varphi_i = (Y_i^{j+1} \wedge Y_{i+1}^j)$.
  \end{itemize}
\item The duo $d_i$ appears twice in $G_2$. We consider two cases: 
  \begin{itemize}
  \item(c) $N_{G_1}[i] = N_{G_1}[i+1]$: in that case, $\varphi_i
    = (\overline{X_{|G_1[i]|} \oplus X_{|G_1[i+1]|}})$ where
    $\oplus$ is the boolean function $XOR$. 
  \item(d) $N_{G_1}[i] \neq N_{G_1}[i+1]$: in that case, $\varphi_i = (X_{|G_1[i]|} \oplus X_{|G_1[i+1]|})$.
  \end{itemize}
\end{enumerate}

Remark that each formula $\varphi_i$ contains two literals. Hence, $(\chi,\Phi)$ is an instance of \PB{Max-2-CSP}. 

\begin{lemma}\label{lemma_CSP1}
  Let $G_1$ and $G_2$ be two balanced genomes such that $\OCC(G_1) =
  2$. Let $(\chi,\Phi)$ be the instance of \PB{Max-2-CSP} obtained by
  $\ALGO{MakeCSP}(G_1,G_2)$. For any integer $k$, if there exists a maximum
  matching $(G_1^M,G_2^M,\mathcal{M})$ of $(G_1,G_2)$ which induces at least $k$
  adjacencies, then there exists an assignment of the variables of
  $\chi$ such that at least $k$ formulas of $\Phi$ are satisfied. 
\end{lemma}

\begin{figure}[hbt]
  \begin{center}
    \includegraphics[width=.4\textwidth]{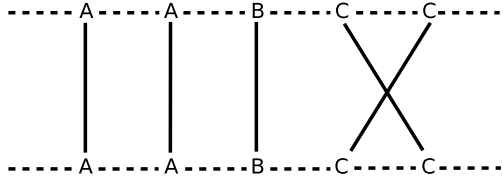}
  \end{center}
  \caption{All possibilities of assignment: $X_A = 1$ (gene $A$
    occurs twice and signed genes are matched in the reading order),
    $X_B = 1$ or $X_B = 0$ (gene $B$ occurs once) and $X_C = 0$
    (gene $C$ occurs twice and signed genes are not matched in the
    reading order). Note that this construction is independent of the
    sign of the genes.} 
  \label{figure_variables}
\end{figure}

\begin{proof}
  Let $G_1$ and $G_2$ be two balanced genomes such that $\OCC(G_1) = 2$
  and let $(\chi,\Phi)$ be the instance of \PB{Max-2-CSP} obtained by
  $\ALGO{MakeCSP}(G_1,G_2)$. Let $k$ be an integer. 

  Suppose there exists a maximum matching $(G_1^M,G_2^M,\mathcal{M})$ of $(G_1,G_2)$
  which induces at least $k$ adjacencies. We construct the following
  assignment of variables of $\chi$. For each gene $g$, we define $X_g =
  1$ if $g$ is not duplicated, else we define $X_g = 1$ iff the
  occurrences of $g$ are matched in the reading order (see Figure
  \ref{figure_variables}). We now show that for each duo which induces
  an adjacency between $G_1^M$ and $G_2^M$, there exists a distinct
  satisfied formula of $\Phi$. Let $d_i = (G_1^M[i],G_1^M[i+1])$, $1
  \leq i \leq n_{G_1} - 1$, be a duo which induces an
  adjacency, and let $d_j = (G_2^M[j],G_2^M[j+1])$ be the related duo on
  $G_2^M$. By construction of $\Phi$, there exists a formula $\varphi_i
  \in \Phi$ which has been previously defined in one of the cases (a),
  (b), (c) or (d) of the definition of \ALGO{MakeCSP}. We claim that, for
  each of these cases, $\varphi_i$ is satisfied: 
  \begin{itemize}
  \item (a) $\varphi_i = (Y_i^j \wedge Y_{i+1}^{j+1})$ and $d_i =
    d_j$. We first prove that literal $Y_i^j$ is true. Three cases are
    possible. \textbf{(i)} The gene $|G_1[i]|$ is not duplicated~; then
    we have defined in our assignment $X_{|G_1[i]|} = 1$. Moreover, we
    have $Y_i^j = X_{|G_1[i]|}$ (since $N_{G_1}[i] = N_{G_2}[j] = 0$),
    hence $Y_i^j$ is true. \textbf{(ii)} The gene $|G_1[i]|$ is
    duplicated and $N_{G_1}[i] = N_{G_2}[j]$~; then, by definition of our
    assignment and since $G_1[i]$ and $G_2[j]$ are matched together in
    the maximum matching $(G_1^M,G_2^M,\mathcal{M})$, we have $X_{|G_1[i]|} = 1$ (we
    match signed genes in the reading order). Moreover, we have $Y_i^j =
    X_{|G_1[i]|}$ which induces that $Y_i^j$ is true.  \textbf{(iii)}
    The gene $|G_1[i]|$ is duplicated and $N_{G_1}[i] \neq N_{G_2}[j]$~;
    then, by definition of our assignment and since $G_1[i]$ and
    $G_2[j]$ are matched together in the maximum matching
    $(G_1^M,G_2^M,\mathcal{M})$, we have $X_{|G_1[i]|} = 0$ (we do not match signed genes
    in the reading order). Moreover, we have in this case $Y_i^j =
    \overline{X_{|G_1[i]|}}$ which induces that $Y_i^j$ is true.  

    In each case, we have proved that $Y_i^j$ is true. We can also prove that
    $Y_{i+1}^{j+1}$ is true, using the same arguments. Hence, we conclude
    that $\varphi_i$ is true. 
  \item(b) $\varphi_i = Y_i^{j+1} \wedge Y_{i+1}^{j}$ and
    $d_i = -d_j$. By similar arguments as in case (a), we can prove
    that $Y_i^{j+1}$ and $Y_{i+1}^{j}$ are true. 
  \item(c) We have $N_{G_1}[i] = N_{G_1}[i+1]$ and the duo $d_i$ appears
    twice in $G_2$ (noted $d_j$ and $d_{j'}$). Since $d_i$ induces an
    adjacency, the duo $d_i$ matches either $d_j$ or $d_{j'}$. In these two
    cases, we have $X_{|G_1[i]|} = X_{|G_1[i+1]|}$ (otherwise $G_1[i]$ and
    $G_1[i+1]$ would not match successive signed genes). Moreover,  $\varphi_i
    = (\overline{X_{|G_1[i]|} \oplus X_{|G_1[i+1]|}})$ and thus,
    $\varphi_i$ is true. 
  \item(d) We have $N_{G_1}[i] \neq N_{G_1}[i+1]$ and the duo $d_i$
    appears twice in $G_2$ (noted $d_j$ and $d_{j'}$). Since $d_i$
    induces an adjacency, the duo $d_i$ matches either $d_j$ or
    $d_{j'}$. In these two cases, we have $X_{|G_1[i]|} \neq
    X_{|G_1[i+1]|}$ (otherwise $G_1[i]$ and $G_1[i+1]$ would not match
    successive signed genes). Moreover, $\varphi_i = (X_{|G_1[i]|} \oplus
    X_{|G_1[i+1]|})$ and thus, $\varphi_i$ is true.  
  \end{itemize}  

  We have constructed a variable assignment of $\chi$ such that, for
  each duo $d_i$ in $G_1^M$ which implies an adjacency, there exists a
  distinct satisfied formula $\varphi_i \in \Phi$. Thus, if there exists
  a maximum matching of $(G_1,G_2)$ which induces at least $k$
  adjacencies, then the corresponding assignment implies at least $k$
  satisfied formulas. \qed  
\end{proof}

\begin{lemma}\label{lemma_CSP2}
  Let $G_1$ and $G_2$ be two balanced genomes such that $\OCC(G_1) =
  2$. Let $(\chi,\Phi)$ be the instance of \PB{Max-2-CSP} obtained by
  $\ALGO{MakeCSP}(G_1,G_2)$. For any integer $k$, if there exists an assignment
  of $\chi$ such that at least $k$ formulas of $\Phi$ are satisfied,
  then there exists a maximum matching $(G_1^M,G_2^M,\mathcal{M})$ of $(G_1,G_2)$
  which induces at least $k$ adjacencies. 
\end{lemma}

\begin{proof}
  Let $G_1$ and $G_2$ be two balanced genomes such that $\OCC(G_1) = 2$
  and let $(\chi,\Phi)$ be the instance of \PB{Max-2-CSP} obtained by
  $\ALGO{MakeCSP}(G_1,G_2)$. Let $k$ be an integer. 

  Suppose there exists an assignment of $\chi$ such that at least $k$
  formulas $\varphi_i \in \Phi$ are satisfied. We create the following
  maximum matching $(G_1^M,G_2^M,\mathcal{M})$ of $(G_1,G_2)$. For each variable
  $X_g$ such that the gene $g$ is duplicated, we match the occurrences
  of $g$ in the reading order if $X_g = 1$ (such as gene $A$ in Figure
  \ref{figure_variables}). If we have $X_g = 0$, we match the first
  occurrence of $g$ on $G_1$ with the second one on $G_2$ and the second
  occurrence of $g$ on $G_1$ with the first one on $G_2$ (such as gene
  $C$ in Figure \ref{figure_variables}). Then, we match signed genes which are
  not duplicated. Now, we prove that each satisfied formula $\varphi_i
  \in \Phi$ induces a distinct adjacency for $(G_1^M,G_2^M,\mathcal{M})$. Let
  $\varphi_i \in \Phi$ be a satisfied formula which is defined in one of
  the cases (a), (b), (c) or (d) of the definition of \ALGO{MakeCSP}: 

  \begin{itemize}
  \item(a) We have $\varphi_i = (Y_i^j \wedge Y_{i+1}^{j+1})$ and the
    duos $d_i=(G_1[i],G_1[i+1])$ and $d_j=(G_2[j],G_2[j+1])$ are
    identical.

    Here, we must prove that $d_i$ and $d_j$ are matched together
    in $(G_1^M,G_2^M,\mathcal{M})$ and thus induce an adjacency. First, we
    show that signed genes $G_1[i]$ and $G_2[j]$ are matched
    together in $(G_1^M,G_2^M,\mathcal{M})$. Since $\varphi_i$ is satisfied,
    we have $Y_i^j=1$. We must dissociate three cases:
    \textbf{(i)} the gene $|G_1[i]|$ is not duplicated: in that
    case, the signed
    gene $G_1[i]$ can be matched only with  $G_2[j]$.
    \textbf{(ii)} The gene $|G_1[i]|$ is duplicated and we have
    $N_{G_1}[i] = N_{G_2}[j]$.  In that case, we have defined
    $Y_i^j = X_{|G_1[i]|}$ which implies $X_{|G_1[i]|}=1$. Thus,
    since $N_{G_1}[i] = N_{G_2}[j]$, the signed genes $G_1[i]$ and
    $G_2[j]$ are matched together. \textbf{(iii)} The gene
    $|G_1[i]|$ is duplicated and we have $N_{G_1}[i] \neq
    N_{G_2}[j]$.  In that case, we have defined $Y_i^j =
    \overline{X_{|G_1[i]|}}$ which implies $X_{|G_1[i]|}=0$. Thus,
    since $N_{G_1}[i] \neq N_{G_2}[j]$, the signed genes $G_1[i]$
    and $G_2[j]$ are matched together. For each case, the signed
    genes $G_1[i]$ and $G_2[j]$ are matched together. We can
    conclude in the same way that $G_1[i+1]$ and $G_2[j+1]$ are
    also matched together, which implies that $d_i$ induces an
    adjacency. 
  \item(b) We have $\varphi_i = (Y_i^{j+1} \wedge Y_{i+1}^j) = 1$ and
    the duos $d_i=(G_1[i],G_1[i+1])$ and $d_j=(G_2[j],G_2[j+1])$ are
    reversed.
    
    We can use the same reasoning  used in case (a) to prove that $d_i$
    induces an adjacency. 
  \item(c) The duo $d_i$ appears twice in $G_2$ (noted $d_j$ and
    $d_{j'}$). We have $\varphi_i = (
    \overline{X_{|G_1[i]|} \oplus X_{|G_1[i+1]|}})$ and
    $N_{G_1}[i] = N_{G_1}[i+1]$.

    Since $\varphi_i$ is true, we have $X_{|G_1[i]|} =
    X_{|G_1[i+1]|}$ which implies by construction of the
    maximum matching that $d_i$ matches $d_j$ or
    $d_{j'}$. 
  \item(d) The duo $d_i$ appears twice in $G_2$ (noted $d_j$ and
    $d_{j'}$). We have $\varphi_i = (X_{|G_1[i]|} \oplus
    X_{|G_1[i+1]|})$ and $N_{G_1}[i] \neq
    N_{G_1}[i+1]$. 
    Since $\varphi_i$ is true, we have $X_{|G_1[i]|} \neq
    X_{|G_1[i+1]|}$ which implies by construction of the
    maximum matching that $d_i$ matches $d_j$ or
    $d_{j'}$. 
  \end{itemize}
  Consequently, for each satisfied formula, there exists a distinct
  adjacency between $G_1^M$ and $G_2^M$. Thus, if there exists an
  assignment of $\chi$ which implies at least $k$ satisfied formulas of
  $\Phi$, then there exists a maximum matching of $(G_1,G_2)$ which
  implies at least $k$ adjacencies. \qed  
\end{proof}

Lemmas \ref{lemma_CSP1} and \ref{lemma_CSP2} prove that any
$\alpha$-approximation for \PB{Max-2-CSP} implies an
$\alpha$-approximation for \PB{Max-2-Adj}. In \cite{approx_CSP}, an
approximation algorithm is given for \PB{Max-2-CSP}, whose approximation ratio
is equal to $\frac{1}{0.874} \leq 1.1442$. Thus, we have the following theorem.  

\begin{theorem}
  \label{approx-max2adj}
  \PB{Max-2-Adj} is $1.1442$-approximable.
\end{theorem}

\subsection{A $3$-approximation for \PB{Max-3-Adj}}
Now, we present a $3$-approximation for \PB{Max-3-Adj} by using the
\PB{Maximum Independent Set} problem defined as follows: 

\medskip
\mabox{\textbf{Problem:} \PB{Max-Independent-Set}\\
  \textbf{Input:} A graph $G=(V,E)$. \\
  \textbf{Solution:} An independent set of $G$ (i.e. a subset
  $V'$ of $V$ such that no two vertices in $V'$ are joined by an
  edge in $E$). \\ 
  \textbf{Measure:} The cardinality of $V'$.
}
\medskip

\begin{figure}[hbt]
  \begin{center}
    \includegraphics[width=.45\textwidth]{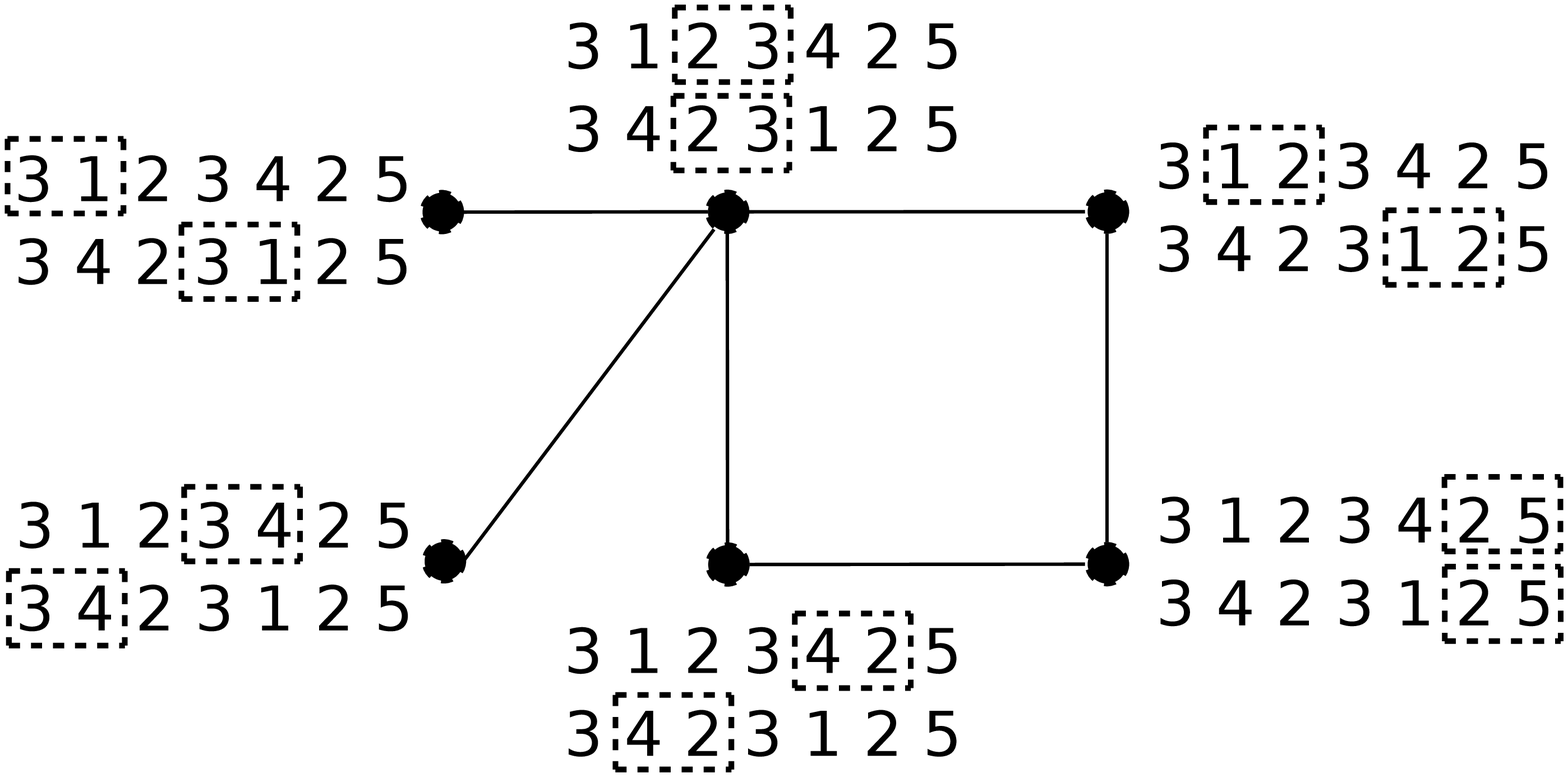}
  \end{center}
  \caption{The conflict graph obtained by $\ALGO{MakeMIS}(G_1,G_2)$
    where $G_1= +3+1+2+3+4+2+5$ and $G_2= +3+4+2+3+1+2+5$ (for
    sake of readability, positive signs are not displayed).} 
  \label{figure_graph_MIS}
\end{figure}

In \cite{breakpoint_approximation_kolmann}, Goldstein \emph{et al}. used
\PB{Max-Independent-Set} to approximate the Minimum Common String Partition problem by
creating a \emph{conflict graph}. We construct in the same way an
instance of \PB{Max-Independent-Set} where a vertex represents a possible adjacency and
where an edge represents a conflict between two adjacencies. We define
$\ALGO{MakeMIS}$ to be the following transformation which associates to two
balanced genomes $G_1$ and $G_2$ an instance of \PB{Max-Independent-Set}. We construct a
vertex for each duo match, and then we create an edge between two
vertices when they are in conflict, i.e. when two matches are
incompatible. Figure \ref{figure_graph_MIS} illustrates the graph
obtained by $\ALGO{MakeMIS}(G_1,G_2)$ where $G_1= +3+1+2+3+4+2+5$ and $G_2=
+3+4+2+3+1+2+5$.  

In order to prove that there exists a $3$-approximation for \PB{Max-3-Adj}, we give the following intermediate lemmas. 

\begin{lemma}\label{lemma_MIS_1}
  Let $G_1$ and $G_2$ be two balanced genomes and let $G$ be the graph
  obtained by \linebreak $\ALGO{MakeMIS}(G_1,G_2)$. For any integer $k$, there exists an
  independent set $V'$ of $G$ such that $|V'| \geq k$ iff there
  exists a maximum matching $(G_1^M,G_2^M,\mathcal{M})$ of $(G_1,G_2)$ which induces
  at least $k$ adjacencies.  
\end{lemma}

\begin{proof}
  Let $G_1$ and $G_2$ be two balanced genomes and let $G$ be the graph
  obtained by \linebreak $\ALGO{MakeMIS}(G_1,G_2)$. Let $k$ be an integer. 

  $(\Rightarrow$) Suppose there exists an independent set $V'$ of $G$
  such that $|V'| \geq k$. We construct a matching $(G_1^M,G_2^M,\mathcal{M})$
  of $(G_1,G_2)$ as follows: first, for each vertex of $V'$, we match
  together the two corresponding duos, thus inducing one adjacency
  (called a \emph{definite} adjacency). By construction of $G$, this
  operation is possible. Indeed, two vertices which are not connected in
  $G$ imply two compatible adjacencies. Then, we match arbitrarily the
  unmatched genes. This operation cannot break any definite
  adjacency. Finally, we obtain a maximum matching $(G_1^M,G_2^M,\mathcal{M})$ which
  induces at least $|V'|$ adjacencies, and consequently at least $k$
  adjacencies. 

  ($\Leftarrow$) Suppose there exists a maximum matching $(G_1^M,G_2^M,\mathcal{M})$
  of $(G_1,G_2)$ which induces at least $k$ adjacencies. We construct a
  set $V'$ by taking each vertex which represents a duo match between
  $G_1^M$ and $G_2^M$. By construction of $G$, $V'$ is an independent
  set (no pair of adjacencies can create a conflict), and then we have
  $|V'| \geq k$. \qed  
\end{proof}

\begin{lemma}\label{lemma_degree}
  Let $G_1$ and $G_2$  be two balanced genomes such that $\OCC(G_1) =
  k$. The maximum degree $\Delta$ of the graph $G$ obtained by
  $\ALGO{MakeMIS}(G_1,G_2)$ satisfies $\Delta \leq 6(k-1)$.   
\end{lemma}

\begin{proof}
  Let $G_1$ and $G_2$ be two balanced genomes such that $\OCC(G_1) = k$
  and let $G$ be the graph obtained by $\ALGO{MakeMIS}(G_1,G_2)$. Consider a
  duo match $m=(d_1,d_2)$ with $d_1=(G_1[i],G_1[i+1])$ and
  $d_2=(G_2[j],G_2[j+1])$ where $1 \leq i \leq n_{G_1} - 1$
  and $1 \leq j \leq n_{G_2} - 1$. We claim that the vertex
  $v_m$ of $G$, which represents the duo match $m$, is connected to at
  most $6(k-1)$ vertices. For this, we list the possible
  duo matches $m'=(d_1',d_2')$ such that the vertex $v_{m'}$ of $G$ which
  represents $m'$ is connected to $v_m$. Remark that if $v_{m'}$ is connected
  to $v_m$ (i.e. $m$ and $m'$ are in conflict), then at least one of the
  duos $d_1'$ and $d_2'$ overlaps, respectively, either $d_1$ or $d_2$. Let $d_1'$
  be a duo in $G_1$ which overlaps $d_1$. First, we list the possible
  duos $d_2'$ such that the duo matches $m=(d_1,d_2)$ and $m'=(d_1',d_2')$
  are in conflict. Remark that $d_1'$ (or $-d_1'$) appears at
  most $k$ times on $G_2$ since a gene can occur at most $k$ times. We
  then distinguish three cases: 

  \begin{itemize}
  \item(a) $d_1' = (G_1[i-1],G_1[i])$: if $d_1'$ (or $-d_1'$)
    appears $k$ times in $G_2$, one of these occurrences is necessary
    $d_2' = (G_2[j-1],G_2[j])$ if $d_1 = d_2$, or $d_2' =
    (G_2[j+1],G_2[j+2])$ if $d_1 = -d_2$. For these two cases,
    the duo matches $m$ and $(d_1',d_2')$ are not in conflict. 
  \item(b) $d_1' = d_1$: if $d_1'$ (or $-d_1'$) appears $k$
    times on $G_2$, one of these occurrences is necessary $d_2$, which
    induces in this case no conflict with $m$. 
  \item(c) $d_1' = (G_1[i+1],G_1[i+2])$: if $d_1'$ (or
    $-d_1'$) appears $k$ times on $G_2$, one of these
    occurrences is necessary $d_2' = (G_2[j+1],G_2[j+2])$ if $d_1 = d_2$,
    or $d_2' = (G_2[j-1],G_2[j])$ if $d_1 = -d_2$. For these
    two cases, the duo matches $m$ and $m'$ are not in
    conflict.  
  \end{itemize} 

  For each case, one of the $k$ possible duos $d_2'$ does not imply
  a conflict between $m$ and $m'$. Thus, for any duo $d_1'$ which
  overlaps $d_1$, there exists at most $k-1$ duos $d_2'$ on $G_2$ such
  that $m$ and $m'$ are in conflict. Using the same arguments, we
  can easily prove that for any duo $d_2'$ which overlaps $d_2$, there
  exists at most $k-1$ duos $d_1'$ on $G_1$ such that $m$ and
  $m'$ are in conflict. Hence, each of the six duos which
  overlaps $d_1$ or $d_2$ implies at most $k-1$ conflicts. Thus, we
  obtain at most $6(k-1)$ vertices which are connected to the vertex
  $v_m$ in the conflict graph. \qed  
\end{proof}

According to Lemma~\ref{lemma_MIS_1}, any $\alpha$-approximation for
\PB{Max-Independent-Set} is thus also an $\alpha$-approximation for
\PB{Max-$k$-Adj}. 
It is proved in \cite{MIS_approximation}
that \PB{Max-Independent-Set} that
is approximable within ratio  $\frac{\Delta+3}{5}$, where $\Delta$ is the
maximum degree of the graph.
Combining this with  Lemma~\ref{lemma_degree}, we obtain the following
result.

\begin{theorem}
  \label{approx-max3adj}
  \PB{Max-$k$-Adj} is $\frac{6k-3}{5}$-approximable.
\end{theorem}

Note that in the case where $k = 2$, we obtain a ratio of $1.8$, which
is not better than the one obtained in
Theorem~\ref{approx-max2adj}. Moreover, we introduce in the next 
section a $4$-approximation in the general case. Hence, the
only interesting case of Theorem~\ref{approx-max3adj} above is when $k = 3$,
inducing a $3$-approximation for \PB{Max-3-Adj}.

\subsection{A $4$-approximation for \PB{Max-Adj}}
In~\cite{approximation_two_interval_pattern}, a $4$-approximation
algorithm for the \PB{Max-Weighted~2-interval~Pattern} problem (\PB{Max-W2IP}) is
given. In the following, we first define \PB{Max-W2IP}, and next we present
how we can relate any instance of \PB{Max-Adj} to an instance of
\PB{Max-W2IP}.

\paragraph{The Maximum Weighted $2$-Interval Pattern problem.}
A 2-\emph{interval} is the union of two disjoint intervals defined
over a single line. For a 2-interval $D=(I,J)$, we always assume that
the interval $I < J$,
\emph{i.e.},
$I$ is completely on the left of $J$ does not overlap $J$. 
We say that two 2-intervals
$D_1=(I_1,J_1)$ and $D_2=(I_2,J_2)$ are \emph{disjoint} if $D_1$ and
$D_2$ have no common point (i.e. $(I_1 \cup J_1)\cap (I_2 \cup J_2) =
\emptyset$). Three possible relations exist between two disjoint
2-intervals: we write 
(1) $D_1 \prec D_2$, if $I_1 < J_1 < I_2 < J_2$, 
(2) $D_1 \sqsubset D_2$, if $I_2 < I_1 < J_1 < J_2$ and
(3) $D_1 \between D_2$, if $I_1 < I_2 < J_1 < J_2$.

We say that a pair of 2-intervals $D_1$ and $D_2$ is $R$-$comparable$
for some $R \in \{\prec,\sqsubset,\between\}$, if either $(D_1,D_2)
\in R$ or $(D_2,D_1) \in R$. A set of 2-intervals $\mathcal{D}$ is
$\mathcal{R}$-comparable for some $\mathcal{R} \subseteq
\{\prec,\sqsubset,\between\}$, $\mathcal{R} \neq \emptyset$, if any
pair of distinct 2-intervals in $\mathcal{D}$ is $R$-comparable for
some $R \in \mathcal{R}$. The non-empty set $\mathcal{R}$ is called a
\emph{model}. 
The \PB{Max-Weighted~2-interval~Pattern} (\PB{Max-W2IP}) problem is
formally defined as follows.

\medskip
\mabox{\textbf{Problem:} \PB{Max-Weighted~2-interval~Pattern} (\PB{Max-W2IP})\\
  \textbf{Input:} A set $\mathcal{D}$ of 2-intervals, a
  model $\mathcal{R}\subseteq
  \{\prec,\sqsubset,\between\}$ with $\mathcal{R} \neq \emptyset$,
  and a weight function $\omega: \mathcal{D} \to \mathbb{N}^+$.\\ 
  \textbf{Solution:} An $\mathcal{R}$-comparable subset $D'$ of
  $\mathcal{D}$.\\ 
  \textbf{Measure:} The weight of $D'$.
}

\paragraph{Transformation.} 
We first describe how to transform any instance $(G_1, G_2)$ of
\PB{Max-Adj} into an instance, referred hereafter as
$\ALGO{Make2I}(G_1, G_2) = (\mathcal{D}, \mathcal{R}, \omega)$,
of \PB{Max-W2IP}.
We need a new definition.
Let $G_1$ and $G_2$ be two balanced genomes. 
An interval $I_1$ of $G_1$ and an interval $I_2$
of $G_2$, both of size at least $2$, are said to be \emph{identical}
if they correspond to the 
same string up to a complete reversal, where a reversal also changes
all the signs in the string.
Clearly, two identical intervals have the same length.

\begin{figure}[hbt]
  \begin{center}
    \includegraphics[width=.4\textwidth]{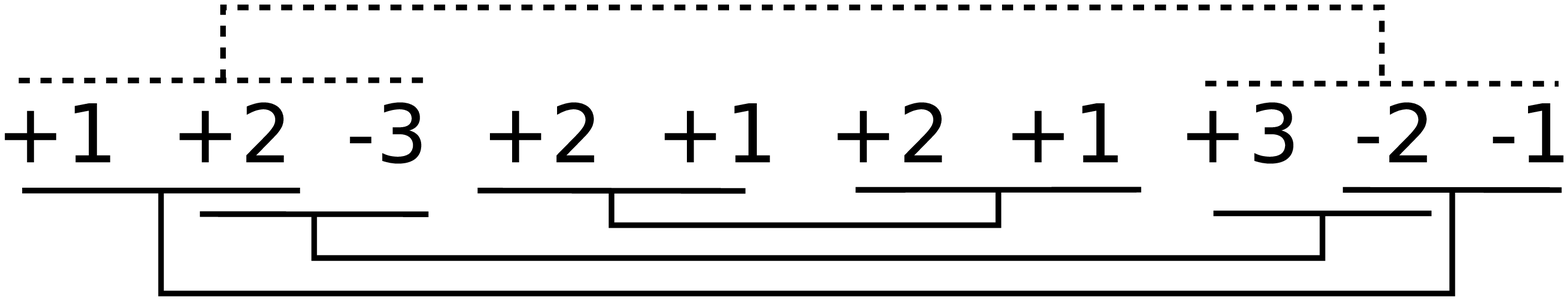}
  \end{center}
  \caption{2-intervals induced by genomes 
    $G_1 = +1~+2~-3~+2~+1$ and $G_2 = +2~+1~+3~-2~-1$. 
    For readability, singleton intervals are not drawn. The dotted
    2-interval is of weight 2, while all other 2-intervals are of
    weight 1.}  
  \label{adjacences_two_intervals}
\end{figure}

The weighted 2-interval set $\mathcal{D}$ is obtained as
follows. 
We first concatenate $G_1$ and $G_2$, 
and
for any pair $(I_1, I_2)$ of identical intervals 
($I_1$ is an interval of $G_1$ and $I_2$ is an interval of $G_2$), 
we construct the 2-interval $D = (I_1, I_2)$ of weight 
$\omega(D) = |I_1| - 1$ ($= |I_2| - 1$)
and add it to $\mathcal{D}$.
Notice that, since identical intervals have length at least $2$, each
2-interval of $\mathcal{D}$ has weight at least $1$.
Figure \ref{adjacences_two_intervals} gives an example of such a
construction. 
Observe that, by construction, no two 2-intervals of 
$\mathcal{D}$ are $\{\prec\}$-comparable.
The construction of the instance of \PB{Max-W2IP} is complete by
setting
$\mathcal{R} = \{\prec, \sqsubset, \between\}$,
\emph{i.e.}, we are looking for disjoint 2-intervals, no matter what
the relation between any two disjoint 2-interval is.
Therefore, for sake of abbreviation, we shall denote the corresponding
instance simply as 
$\ALGO{Make2I}(G_1, G_2) = (\mathcal{D}, \omega)$ and forget about
the model. 

We now describe how to transform any solution of \PB{Max-W2IP} into a
solution of \PB{Max-Adj}. 
Let $G_1$ and $G_2$ be two balanced genomes and 
$\ALGO{Make2I}(G_1, G_2) = (\mathcal{D}, \omega)$. 
Furthermore, let $\mathcal{S} \subseteq \mathcal{D}$ be a set of
disjoint 2-intervals, \emph{i.e.} a solution for \PB{Max-W2IP} for model
the $\{\prec,\sqsubset,\between\}$ for 
the instance $(\mathcal{D}, \omega)$. 

We write $\ALGO{Max-W2IP\_to\_Adj}(\mathcal{S})$ for
the transformation of $\mathcal{S}$ into a maximum matching
$(G_1^M, G_2^M, \mathcal{M})$ of $(G_1,G_2)$ defined as follows. 
First, for each
2-interval $D = (I_1,I_2)$ of $\mathcal{S}$, 
we match the signed genes of $I_1$ and $I_2$
in the natural way~; then, in order to achieve a maximum matching
(since each signed gene is not necessarily covered by a 2-interval 
in $\mathcal{S}$),
we apply the following greedy algorithm: iteratively, we match,
arbitrarily, two unmatched signed genes $g_1$ and  $g_2$ such that
$|g_1|=|g_2|$ and $g_i$ is a gene of $G_i$ ($i = 1,2$), until
no such pair of signed genes exists. 
After a relabeling of signed genes
according to this matching (denoted $\mathcal{M}$), we obtain a
maximum matching $(G_1^M, G_2^M, \mathcal{M})$ of $(G_1, G_2)$. 

The rationale of this construction stems from two following lemmas.

\begin{lemma}\label{lemme1Approximation}
  Let $G_1$ and $G_2$ be two balanced genomes,
  $\ALGO{Make2I}(G_1,G_2) = (\mathcal{D}, \omega)$ and
  $\mathcal{S}$ be any set of disjoint
  2-intervals of $\mathcal{D}$. 
  If we denote by $W_\mathcal{S}$ the total weight of $\mathcal{S}$,
  then the maximum matching $(G_1^M, G_2^M, \mathcal{M})$ of 
  $(G_1, G_2)$ obtained
  by $\ALGO{Max-W2IP\_to\_Adj}(\mathcal{S})$ induces at least 
  $W_\mathcal{S}$
  adjacencies.   
\end{lemma}

\begin{proof}
  For each 2-interval $D=(I_1, I_2)$ of $\mathcal{S}$,
  we have matched the signed genes of $I_1$ and $I_2$ in the natural
  way. 
  Therefore, for each 2-interval $D = (I_1, I_2)$ of $\mathcal{S}$,
  we obtain $|I_1|-1$ adjacencies in $(G_1^M, G_2^M, \mathcal{M})$
  since $I_1$ and $I_2$ are identical intervals. 
  Since the final greedy part of $\ALGO{Max-W2IP\_to\_Adj}(\mathcal{S})$
  does not delete any adjacency, we have at least
  $W_\mathcal{S}$ adjacencies in $(G_1^M, G_2^M, \mathcal{M})$.
  \qed
\end{proof}

\begin{lemma}\label{lemme2Approximation}
  Let $G_1$ and $G_2$ be two balanced genomes,
  $(G_1^M, G_2^M, \mathcal{M})$ be a maximum matching of $(G_1,G_2)$,
  $\ALGO{Make2I}(G_1,G_2) = (\mathcal{D}, \omega)$ and 
  $W$ be the number of adjacencies induced by 
  $(G_1^M, G_2^M, \mathcal{M})$. 
  Then there exists a subset $\mathcal{S} \subseteq \mathcal{D}$ of
  disjoint 2-intervals of total weight $W$.
\end{lemma}

\begin{proof}
  Denote by $n$ the size of $G_1^M$.
  Consider any factorization
  $G_1^M = s_1 \, s_2 \, \ldots s_p$ such that,
  for each $1 \leq i < p$, $s_i$ and
  $s_{i+1}$ are separated by one breakpoint and no breakpoint
  appears in $s_i$, $1 \leq i \leq p$.
  Therefore, there exists $p-1$ breakpoints between $G_1^M$ and
  $G_2^M$, and hence $n-p$ adjacencies between $G_1^M$ and
  $G_2^M$
  . To each substring $s_i$ of the factorization of $G_1^M$
  corresponds a substring $t_i$ in $G_2^M$ such that $s_i$ and $t_i$
  are identical. 
  Moreover, each substring $s_i$ of size $l_i$, $1 \leq i \leq p$,
  contains $l_i -1$ adjacencies. 
  We construct the 2-interval set $\mathcal{S}$ as the union
  of $D_i = (\hat{s_i}, \hat{t_i})$, $1 \leq i \leq p$, where
  $\hat{s_i}$ (resp. $\hat{t_i}$) is the interval obtained from $s_i$
  (resp. $t_i$). The factorization of $G_1^M$
  implies that the constructed 2-intervals
  are disjoint, and hence the total weight of $\mathcal{S}$ is 
  $\sum_{i=1}^p{(l_i-1)}=\sum_{i=1}^p{l_i}-\sum_{i=1}^p{1} = n - p = W$.
  \qed   
\end{proof}

We now describe Algorithm \ALGO{ApproxAdj} and then prove it to be
a $4$-approximation algorithm for \PB{Max-Adj}.

\begin{algorithm}
  \caption{\ALGO{ApproxAdj}}
  \begin{algorithmic}
    \REQUIRE Two balanced genomes $G_1$ and $G_2$.
    \ENSURE A maximum matching $(G_1^M, G_2^M, \mathcal{M})$ of
    $(G_1,G_2)$.  
    \STATE \begin{itemize}
    \item Let $\ALGO{Make2I}(G_1,G_2) = (\mathcal{D}, \omega)$. 
    \item Invoke the $4$-approximation algorithm of Crochemore
      \emph{et al}.~\cite{approximation_two_interval_pattern} to
      obtain a set of disjoint 2-intervals 
      $\mathcal{S} \subseteq \mathcal{D}$.
    \item Construct the maximal matching 
      $(G_1^M,G_2^M,\mathcal{M}) = \ALGO{Max-W2IP\_to\_Adj}(\mathcal{S})$.
    \end{itemize}
  \end{algorithmic}
\end{algorithm}

\begin{theorem}\label{theoremeApproximation}
  Algorithm \ALGO{ApproxAdj} is a $4$-approximation algorithm for
  \PB{Max-Adj}. 
\end{theorem}

\begin{proof}
  According to Lemmas~\ref{lemme1Approximation} and
  \ref{lemme2Approximation}, there exists a maximum matching
  $(G_1^M,G_2^M,\mathcal{M})$ of $(G_1, G_2)$ that induces $W$
  adjacencies iff
  there exists a subset of disjoint 2-intervals
  $\mathcal{S} \subseteq \mathcal{D}$ with total weight $W$.
  Therefore, any approximation ratio for \PB{Max-W2IP} implies the
  same approximation ratio for \PB{Max-Adj}.
  In~\cite{approximation_two_interval_pattern}, a
  $4$-approximation algorithm is proposed for \PB{Max-W2IP}.
  Hence, Algorithm \ALGO{ApproxAdj} is a
  $4$-approximation algorithm for \PB{Max-Adj}.  
  \qed 
\end{proof}

\section{Conclusions and future work}
In this paper, we have first given new approximation complexity
results for several optimization problems in genomic rearrangement. We
focused on conserved intervals, common intervals and breakpoints, and we
took into account the presence of duplicates. We restricted our proofs
to cases where one genome contains no duplicates and the other
contains no more than two occurrences of each gene. With this
assumption, we proved that the problems consisting in computing an
exemplarization (resp. an intermediate matching, a maximum matching)
optimizing any of the three above mentioned measures is \APXhard, thus
extending the results
of~\cite{Complexite_exemplarisation,NP_completude_EComI_MComI,chen_inproceding_approximation_breakpoint}.
In a second part of the paper, we have focused on the \PB{ZEBD}
(resp. \PB{ZIBD}, \PB{ZMBD}) problems, where the question is whether there exists an
exemplarization (resp. intermediate matching, maximum matching) that
induces zero breakpoint. We have extended a
result from~\cite{chen_inproceding_approximation_breakpoint} by
showing that \PB{ZEBD} is \NPC even for instances 
of type $(2,k)$, where $k$ is unbounded. We also have noted that
\PB{ZEBD} and \PB{ZIBD} are equivalent problems, and shown that 
\PB{ZMBD} is in \Pclass.
Finally, we gave several approximation algorithms
for computing the maximum number of adjacencies of two balanced genomes under
the \emph{maximum matching} model. The approximation ratios we get
are 1.1442 for instances of type $(2,2)$, 3 for instances of type
$(3,3)$ and 4 in the general case. Concerning the latter result, we
note that the approximation ratio we obtain is constant, even when the
number of occurrences in genomes is unbounded.

However, several problems remain unsolved. In particular, concerning
approximation algorithms, virtually nothing is known (i)~in the case of
unbalanced genomes and (ii)~in the exemplar and intermediate
models. Indeed, all the existing results (see for
instance~\cite{breakpoint_approximation_kolmann,inproceeding_kolman_approximation_breakpoint}
for the number of breakpoints), including ours, focus on the maximum
matching problem for balanced genomes, which 
implies that no gene is deleted from genomes $G_1$ and $G_2$. Now, if
we allow genes 
to be deleted, the problem seems much more difficult to tackle.

Finally, we would like to recall the following open problem
from~\cite{chen_inproceeding_approximation_intervalles_conserves}:
what is the 
complexity of \PB{ZEBD} for instances of type $(2,2)$~?



\bibliographystyle{plain}
\bibliography{bibtex}

\end{document}